\documentclass[amsmath,amssymb,prl,superscriptaddress,showpacs,longbibliography,twocolumn]{revtex4-1}
\usepackage{graphicx}
\usepackage{dcolumn}
\usepackage[colorlinks=true,linkcolor=blue,citecolor=blue,urlcolor=blue]{hyperref}
\usepackage{stmaryrd}
\usepackage{esvect}
\usepackage{natbib}
\usepackage[usenames,dvipsnames]{xcolor}
\usepackage{soul}
\usepackage[separate-uncertainty = true]{siunitx} %new version combining siunits and sistyle package

\newcommand{\vmod}{V_{\mathrm{mod}}}
\newcommand{\vstab}{V_{\mathrm{stab}}}
\newcommand{\istab}{I_{\mathrm{stab}}}

\newcommand{\didu}{$\mathrm{d}I/\mathrm{d}V$ spectrum}
\newcommand{\didus}{$\mathrm{d}I/\mathrm{d}V$ spectra }
\newcommand{\didusig}{$\mathrm{d}I/\mathrm{d}V$ }

\begin{document}

%\title{Effect of substrate spin-orbit coupling and magnetic anisotropy on the topology of the Shiba band structure}
\title{Coexistence of topologically trivial and non-trivial Yu--Shiba--Rusinov bands in magnetic atomic chains on a superconductor}

\author{Bendeg\'uz Ny\'ari}
\affiliation{Department of Theoretical Physics, Budapest University of Technology and Economics, 1111 Budapest, Hungary}
\affiliation{HUN-REN-BME Condensed Matter Research Group, Budapest University of Technology and Economics, 1111 Budapest, Hungary}
\author{Philip Beck}
%\affiliation{Department of Physics, University of Hamburg, Hamburg, Germany}
\affiliation{Institute of Nanostructure and Solid State Physics, University of Hamburg, 20355 Hamburg, Germany}
\author{Andr\'as L\'aszl\'offy}\email{laszloffy.andras@wigner.hun-ren.hu}
\affiliation{Department of Theoretical Solid State Physics, HUN-REN Wigner Research Centre for Physics, 1525 Budapest, Hungary}
\author{Lucas Schneider}
%\affiliation{Department of Physics, University of Hamburg, Hamburg, Germany}
\affiliation{Department of Physics, University of California, Berkeley, 94720 California, United States}
%\affiliation{Materials Sciences Division, Lawrence Berkeley National Laboratory, Berkeley, 94720 California, United States}
\affiliation{Institute of Nanostructure and Solid State Physics, University of Hamburg, 20355 Hamburg, Germany}
\author{Krisztián Palotás}
\affiliation{Department of Theoretical Solid State Physics, HUN-REN Wigner Research Centre for Physics, 1525 Budapest, Hungary}
%\affiliation{Department of Theoretical Physics, Budapest University of Technology and Economics, 1111 Budapest, Hungary}
%\affiliation{HUN-REN-SZTE Reaction Kinetics and Surface Chemistry Research Group, University of Szeged, 6720 Szeged, Hungary}
\author{László Szunyogh}
\affiliation{Department of Theoretical Physics, Budapest University of Technology and Economics, 1111 Budapest, Hungary}
\affiliation{HUN-REN-BME Condensed Matter Research Group, Budapest University of Technology and Economics, 1111 Budapest, Hungary}
\author{Roland Wiesendanger}
%\affiliation{Department of Physics, University of Hamburg, Hamburg, Germany}
\affiliation{Institute of Nanostructure and Solid State Physics, University of Hamburg, 20355 Hamburg, Germany}
\author{Jens Wiebe}
%\affiliation{Department of Physics, University of Hamburg, Hamburg, Germany}
\affiliation{Institute of Nanostructure and Solid State Physics, University of Hamburg, 20355 Hamburg, Germany}
\author{Bal\'azs \'Ujfalussy}\email{ujfalussy.balazs@wigner.hun-ren.hu}
\affiliation{Department of Theoretical Solid State Physics, HUN-REN Wigner Research Centre for Physics, 1525 Budapest, Hungary}
\author{Levente R\'ozsa}
\affiliation{Department of Theoretical Solid State Physics, HUN-REN Wigner Research Centre for Physics, 1525 Budapest, Hungary}
\affiliation{Department of Theoretical Physics, Budapest University of Technology and Economics, 1111 Budapest, Hungary}

\date{\today}

\begin{abstract}
Majorana zero modes (MZMs) have been proposed as a promising basis for Majorana qubits offering great potential for topological quantum computation. Such modes may form at the ends of a magnetic atomic chain on a superconductor. Typically only a single MZM may be present at one end of the chain, but symmetry may protect multiple MZMs at the same end. %Majorana zero modes (MZMs) may form at the ends of a magnetic atomic chain on a superconductor via the opening of a topologically non-trivial gap in the band of Yu--Shiba--Rusinov (YSR) bound states. %, but their experimental observation remains controversial.
Here, we study the topological properties of Yu--Shiba--Rusinov (YSR) bands of excitations in Mn chains constructed on a Nb(110) and on a Ta(110) substrate using first-principles calculations and scanning tunneling microscopy and spectroscopy experiments. We demonstrate that even and odd YSR states with respect to mirroring on the symmetry plane containing the chain have different dispersions, and both of them may give rise to MZMs separately. Although the spin--orbit coupling leads to a hybridization between the bands, multiple MZMs may still exist due to the mirror symmetry. These findings highlight the influence of symmetries on interpreting the spectroscopic signatures of candidates for MZMs.
%Here, we demonstrate the emergence of multiple Yu--Shiba--Rusinov bands with distinct topological properties in Mn chains constructed on a Nb(110) or a Ta(110) substrate using first-principles calculations and scanning tunneling microscopy and spectroscopy experiments. On the Nb substrate, states which are even under mirroring on the plane including the chain axis form a minigap devoid of in-gap states that is topologically trivial, while low-energy states oscillating in energy are observed for the odd states with signatures consistent with that of precursors of Majorana zero modes. On the Ta substrate, delocalized low-energy states are observed both for even and odd states, but rotating the magnetization from along the chain axis to the out-of-plane direction in the calculations leads to a minigap opening in the even states, with precursors of Majorana zero modes localized at the ends of the chains. These findings highlight the influence of multiple Yu--Shiba--Rusinov bands and their spin--orbit-coupling-induced hybridization on interpreting the spectroscopic signatures of candidates for Majorana zero modes.
\end{abstract}

\maketitle

%\emph{Introduction.}
\section{Introduction}

Majorana zero modes (MZMs) have attracted considerable research attention recently because of their proposed applications in topological quantum computing~\cite{Nayak2008,Beenakker2020}. %,Microsoft2025}. 
A pair of MZMs manifests as a fermionic excitation localized at the two ends of a magnetic chain or wire in proximity to a superconductor, which is energetically placed at the Fermi level, or zero energy, inside the superconducting gap~\cite{kitaevchain,Lutchyn2010,oreg2010helical}. Generally, only a single MZM is allowed to exist at one chain end, since pairs of MZMs may hybridize and move away from zero energy. Mathematically this can be described as the MZMs being protected by the particle-hole constraint of superconducting excitations, leading to a $\mathbb{Z}_{2}$ topological classification in symmetry class D. While the magnetism required for the emergence of MZMs breaks the time-reversal symmetry, %of the system, 
it has been suggested that an effective time-reversal symmetry may be restored if the chain is located in a mirror plane of the system~\cite{Fang2014}. This additional symmetry places the system in the BDI symmetry class with an integer $\mathbb{Z}$ topological invariant, i.e., multiple MZMs may coexist at the same chain end~\cite{Tewari2012}. It has been proposed that multiple MZMs are particularly likely to emerge if multiple electronic bands are located in the vicinity of the Fermi level~\cite{li2014topological}. %Theoretically they may be realized in a one-dimensional single-band $p$-wave superconductor~\cite{kitaevchain}, where a pair of MZMs manifests as a fermionic excitation localized at the two ends of the system, energetically placed at the Fermi level, or zero energy, inside the superconducting gap. It has been suggested that MZMs could also be realized in the presence of spin--orbit coupling (SOC) in a one-dimensional ferromagnetic system coupled to a bulk $s$-wave superconductor~\cite{Lutchyn2010,oreg2010helical}, which are more abundant in nature than $p$-wave superconductors. Due to the difficulty of performing the topological quantum operations that involve moving the MZMs around, experimental studies so far have concentrated on identifying their spectroscopic signatures, namely the observation of zero-energy peaks at the ends of the one-dimensional wires or atomic chains~\cite{Mourik2012,Das2012,NadjPerge2014,Ruby2015,Pawlak2016,Jeon2017}. These experiments were often accompanied by theoretical calculations based on the above-mentioned one-dimensional models. However, the reported observations of MZMs have attracted controversy since these spectroscopic signatures are not unique to MZMs, they may also occur in topologically trivial systems. From the experimental side, the atom-by-atom design of magnetic chains enabled by manipulation induced by the tip of a scanning tunneling microscope (STM) made it possible to obtain more detailed scanning tunneling spectroscopic (STS) data which could support or reject the topologically non-trivial origin of the observed zero-energy end states~\cite{Kim2018,Schneider2021a,Schneider2021b,Liebhaber2022,Kster2022}.

Experimental studies on MZMs have concentrated on the observation of zero-energy peaks at the ends of the one-dimensional wires or atomic chains using spectroscopic methods~\cite{Mourik2012,Das2012,NadjPerge2014,Ruby2015,Pawlak2016,Jeon2017,LoConte2025}. However, these spectroscopic signatures may also occur in topologically trivial systems where no MZMs are present, and they do not enable to distinguish between a single or possibly more MZMs protected by the symmetry. Theoretical tight-binding simulations based on material-specific parameters obtained from first-principles calculations~\cite{NadjPerge2014,li2014topological,Kim2018,Schneider2020,Kster2022,Crawford2022} have taken multiple bands into account. However, the large difference between the electronic bandwidth and the superconducting energy gap makes it difficult to estimate the error of the parameters %introduced by the approximations involved in mapping to 
in the tight-binding models, that influences which of these bands are relevant for the formation of MZMs. %while low relative changes in the normal-state hopping parameters may induce large differences on the three orders of magnitude lower energy scale of superconducting models.
This problem is circumvented when superconductivity and the in-gap states are described directly in the first-principles calculations~\cite{csire2018relativistic,Nyari2021,laszloffy2023YSR,Kyungwha2023,nyari2023,laszloffy2023,Ruessmann2022}.

Signatures of MZMs have been recently studied in magnetic chains designed atom by atom via manipulation by the tip of a scanning tunneling microscope (STM)~\cite{Kim2018,Schneider2021a,Schneider2021b,Liebhaber2022,Kster2022,LoConte2025}. In these systems, MZMs emerge from the hybridization of Yu--Shiba--Rusinov (YSR) states~\cite{yu1965bound,shiba1968classical,rusinov1969superconductivity} formed around single magnetic adatoms~\cite{Yazdani1997,Ruby2016,Choi2017,Kezilebieke2018,Schneider2019,Beck2021}. The high degree of control over the structure enables following the evolution of the YSR bands with the chain length, which made it possible to exclude topologically non-trivial contributions to the zero-energy peaks~\cite{Schneider2021b,Schneider2023}. It also enabled the observation of two different types of YSR bands in nearest-neighbor Mn chains built along the $[001]$ direction on the Nb(110) surface~\cite{Schneider2021a}. One band has a high intensity along the center of the chain, and a large minigap is opened in it by the spin--orbit coupling (SOC), but within this minigap no localized end states indicating MZMs could be observed. The other band has enhanced intensities on the two sides of the chain, and such states can be observed arbitrarily close to the Fermi level for the investigated wide array of chain lengths. Theoretical calculations~\cite{Crawford2022} predicted that these second types of states could turn into MZMs with side features if a stronger SOC would be considered, but the role of symmetry or the connection to the states with the high intensity along the center of the chain was not explored. %It remains to be explained why the SOC in the same material system opens a large gap in one YSR band and no observable gap in the other band.

Here, we investigate the possible formation of MZMs in atomic chains with multiple YSR bands transforming differently under the mirror symmetry. We study nearest-neighbor Mn chains along the $[001]$ direction on Nb(110) and Ta(110) surfaces by first-principles calculations based on the screened Korringa--Kohn--Rostoker (SKKR) method~\cite{Szunyogh1995}, and compare these to scanning tunneling spectroscopy (STS) measurements. We identify two types of YSR states which are even and odd with respect to the mirror plane going through the chain axis, similarly to previous experimental observations~\cite{Schneider2021a}. Our calculations suggest that the large minigap in the even band is opened by the SOC, but it does not host MZMs. The low-energy states found in the odd band approach zero energy for higher values of the SOC, consistent with precursors of MZMs. On the Ta substrate, the even band now appears to be topologically non-trivial in the calculations, but the %abundance of low-energy states 
small size of the minigap for the energetically preferred magnetization direction along the chain axis prevents the observation of end states. Changing the magnetization to out of plane in the simulations opens a larger minigap in the even band, within which well-localized end states resembling MZMs are formed.

\section{Results}
\subsection{Mn chains on Nb(110)}
\begin{figure*}[!t]
    \centering
    \includegraphics[width=1\textwidth]{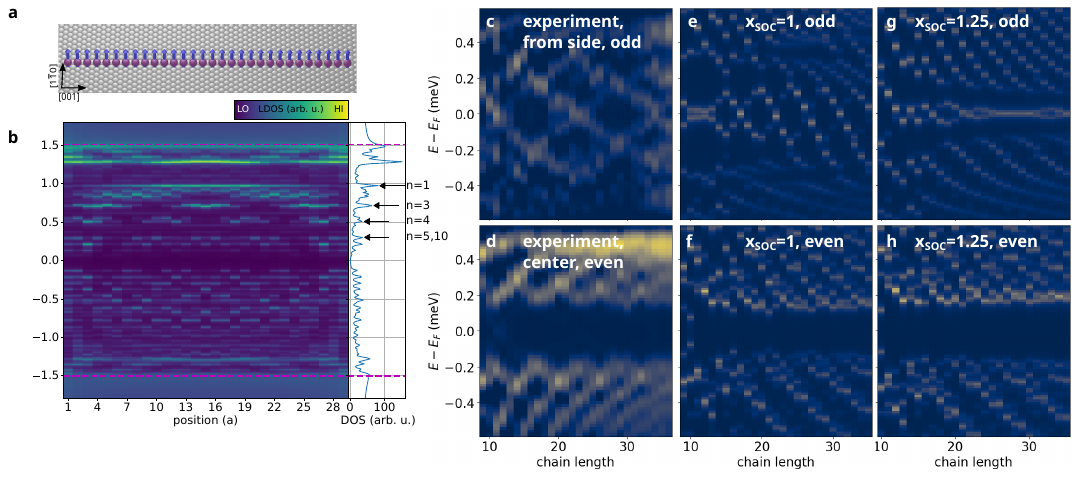}
    \caption{\label{fig:nb-dos}
    Spectra of Mn chains along the [001] direction on Nb(110) in real space. \textbf{a}, Visualization of the %1$a$--
     Mn$_{30}$ chain. The Nb and Mn atoms are shown with gray and purple spheres, respectively, while the out-of-plane magnetization direction is displayed with blue arrows. \textbf{b}, Electron component of the local density of states (LDOS, given in arbitrary units) extracted from the vacuum above the axis ($y$) of the Mn$_{30}$ chain. The LDOS integrated along the chain is shown to the right. Labels $n$ denote the number of observed maxima in the states at energies highlighted by arrows. 
    \textbf{c}, Deconvoluted \didus measured %on chains of length $N$. The spectra were measured 
    %at an end atom of the chain at the side of the atom 
    at the side of one end of all chains in Ref.~\cite{Crawford2022} while additional Mn atoms were attached to the chain's other end, thus 
    %along the side of the chain 
    %in Ref.~\cite{Crawford2022}, 
    being sensitive to the odd states. \textbf{d}, Deconvoluted \didus measured %on chains of length $N$. The spectra were measured 
    and averaged along the center line of the chains %Same as in panel \textbf{c}, with the \didus measured along the center of the chains 
    in Ref.~\cite{Schneider2021a}, being sensitive to the even states. Measurement parameters: $\vstab=\SI{-6}{\milli \volt}$, $\istab=\SI{1}{\nano \ampere}$, $\vmod = \SI{20}{\micro \volt}$.
    \textbf{e}-\textbf{h}, LDOS calculated in the vacuum from one end of the chain as a function of the Mn chain length ranging from 10 to 36, projected to odd or even orbitals as indicated. The scaling factor of the SOC $x_{\textrm{SOC}}$~\cite{SOCscaleHubert} is also given.
    }
\end{figure*}

First, we studied nearest-neighbor (NN) Mn chains on the Nb(110) surface along the [001] direction using first-principles calculations (see Methods), as illustrated in Fig.~\ref{fig:nb-dos}a. 
We will refer to the chains as Mn$_{L}$, where $L$ denotes the number of atoms. The magnetic structure of the chains was found to be ferromagnetic with out-of-plane magnetization, see Supplementary Note 1 and Supplementary Fig.~1, with the ferromagnetic ordering also confirmed by spin-polarized STM measurements~\cite{Schneider2021c}. %Details of the first-principles calculations are given in the Methods section. We determined spin-model parameters from these calculations, and numerically minimized the resulting classical atomistic spin Hamiltonian to find the magnetic ground state, see \hl{Supplementary Note 1 and Supplementary Fig.~1}. The out-of-plane ferromagnetic configuration was energetically the most favorable, in agreement with previous calculations~\cite{Laszloffy2021}.
%The ferromagnetic ordering was also confirmed by spin-polarized STM measurements~\cite{Schneider2021a}, so we used this magnetic configuration for the calculations in the superconducting state. 

Figure~\ref{fig:nb-dos}b shows the local density of states (LDOS) of the Mn$_{30}$ chain as the function of energy relative to the Fermi level and the spatial coordinate along the chain; see Supplementary Movie 1 for other chain lengths. %At the right side of the panel we also plotted the LDOS integrated along the chain. We show the LDOS calculated in the vacuum above the chain, since this is expected to correspond to the differential conductance \didusig measured in STS experiments. 
YSR states appear as peaks in the LDOS inside the superconducting gap $\Delta_{\textrm{Nb}}=1.51$~meV and spatially localized %on or 
in the vicinity of the chain. These YSR states may be characterized based on their spatial profiles using symmetry arguments. Since the chain is built along the [001] direction denoted by $y$, it is located in the $yz$ mirror plane perpendicular to the surface. Combining the mirroring $\sigma_{yz}$ with time reversal $\mathcal{T}$ is a symmetry of the system, since the magnetic moments are located in the mirror plane. This is the effective time-reversal symmetry that can protect multiple MZMs~\cite{Fang2014}. The YSR states on the chain stemming from the atomic states resembling $d_{xy}$ and $d_{xz}$ orbitals~\cite{Beck2021,Schneider2021a,Nyari2021} are odd under mirroring, and have a nodal line in their LDOS along the axis of the chain. The atomic states resembling $d_{z^{2}}$, $d_{x^{2}-y^{2}}$ and $d_{yz}$ orbitals~\cite{Beck2021,Schneider2021a,Nyari2021} give rise to even YSR states on the chain, which typically have a high LDOS along the center of the chain. This difference in intensity between the two types of states makes it possible to distinguish them using STS experiments.

Spectra measured on the side of the chains in Ref.~\cite{Crawford2022} are shown in Fig.~\ref{fig:nb-dos}c as a function of chain length. These measurements were primarily sensitive to the odd states displaying pronounced side features. %These \didus integrated along the chain are shown in Fig.~\ref{fig:nb-dos}c as a function of chain length. 
The lowest-lying state oscillates in energy with the chain length due to the finite-size confinement, and crosses the Fermi level multiple times. For certain chain lengths, this state may be located at $E_{\textrm{F}}$ (see, e.g.,~the Mn$_{34}$ chain), but it is always extended along the whole chain~\cite{Schneider2021a,Crawford2022}, and increasing or decreasing the chain length by a single atom moves it away in energy. These oscillations, although with a shorter period, are reproduced by the calculations in Fig.~\ref{fig:nb-dos}e, showing the LDOS projected to the atomic orbitals which are odd under mirroring. While these features are consistent with precursors of MZMs~\cite{Schneider2021b,Stanescu2013,Albrecht2016,Lee2013} in short chains, longer chains would be necessary to observe MZMs which are localized at the ends and are energetically separated from the YSR bands by a minigap. In the simulations, better localization of the low-energy states may be achieved by scaling the strength of the SOC by a factor of $x_{\textrm{SOC}}=1.25$~\cite{SOCscaleHubert}, as shown in Fig.\ref{fig:nb-dos}g. This opens a minigap of $\Delta_{\textrm{mini},x_{\textrm{SOC}}=1.25,\textrm{odd}}=0.13~\mathrm{meV}$, wherein only a single state can be observed. This is similar to the calculations in Ref.~\cite{Crawford2022} when the SOC was increased in the tight-binding model. The lowest-energy state appears to converge to the Fermi level at chain lengths of around 30 atoms, and its intensity starts to become localized at the two ends; see Supplementary Movie 2 for the real-space distribution of the LDOS. However, the energy of the lowest-lying state %starts to increase 
moves away from the Fermi level again in longer chains.

In the spectra measured along the centers of the chains in Ref.~\cite{Schneider2021a} and shown in Fig.~\ref{fig:nb-dos}d, the even states (denoted by $\alpha$ in Ref.~\cite{Schneider2021a}) have been observed to have a much higher intensity than the odd states (denoted by $\delta$ in Ref.~\cite{Schneider2021a}). %These \didus are shown in Fig.~\ref{fig:nb-dos}d. 
Starting from chain lengths of around 10 atoms, a minigap is fully developed in the energy range of $\pm\Delta_{\textrm{mini,exp}}=\pm 0.18$~meV, inside of which only the odd states may be observed with a faint intensity (cf. Fig.~\ref{fig:nb-dos}c for the same features). Note that the even states also display crossing features similar to %the Fermi-level crossings 
those of the odd states discussed above, but these can only be observed outside $\Delta_{\textrm{mini,exp}}$ in this case, e.g., close to $E-E_{\textrm{F}}=\pm 0.30$~meV. The presence of these crossings above and below $\Delta_{\textrm{mini,exp}}$ indicates that the minigap is most likely opened by the SOC, and YSR states could be observed at every energy in the absence of SOC. %, it was concluded in Ref.~\cite{Schneider2021a} that the gap in the YSR band is of topological origin most likely opened by the SOC. 
Surprisingly, no signature of MZMs or their precursors were observed for the even states. In the corresponding simulations in Fig.~\ref{fig:nb-dos}f, a minigap of $\Delta_{\textrm{mini,calc}}=0.13$~meV opens in the YSR band, also without any low-energy states. The calculations confirm that the SOC is responsible for opening the gap in the even states, since scaling it by a factor of $x_{\textrm{SOC}}=0$ closes the minigap (see Supplementary Note 2 and Supplementary Figure 2), while a scaling factor of $x_{\textrm{SOC}}=1.25$ results in an increase to $\Delta_{\textrm{mini},x_{\textrm{SOC}}=1.25,\textrm{even}}=0.15~\mathrm{meV}$ in Fig.~\ref{fig:nb-dos}h.

The number of MZMs in the chain is expected to be equal to the topological invariant of the YSR bands in Fourier space. Reciprocal-space information is encoded in the periodicity of the LDOS in real space; for example, in Fig.~\ref{fig:nb-dos}b periodic modulations with one, three, and four maxima can be observed at energies $E-E_{\textrm{F}}=0.98$~meV, $E-E_{\textrm{F}}=0.71$~meV, and $E-E_{\textrm{F}}=0.51$~meV, respectively. These periodic modulations may be interpreted as quasiparticle interference (QPI) patterns between excitations of the same energy, but different wave vectors. For example, oppositely propagating waves with wave vectors $k$ and $-k$ produce a QPI with scattering vector $q=k-(-k)=2k$. Consequently, taking the spatial Fourier transforms of the measured \didusig line profiles along the chain or of the calculated LDOS, and combining them for different chain lengths gives information on the %long-wavelength limit of the 
YSR dispersion relation~\cite{Schneider2021a,beck2023search}.

\begin{figure*}[!t]
    \centering
    \includegraphics[width=1\textwidth]{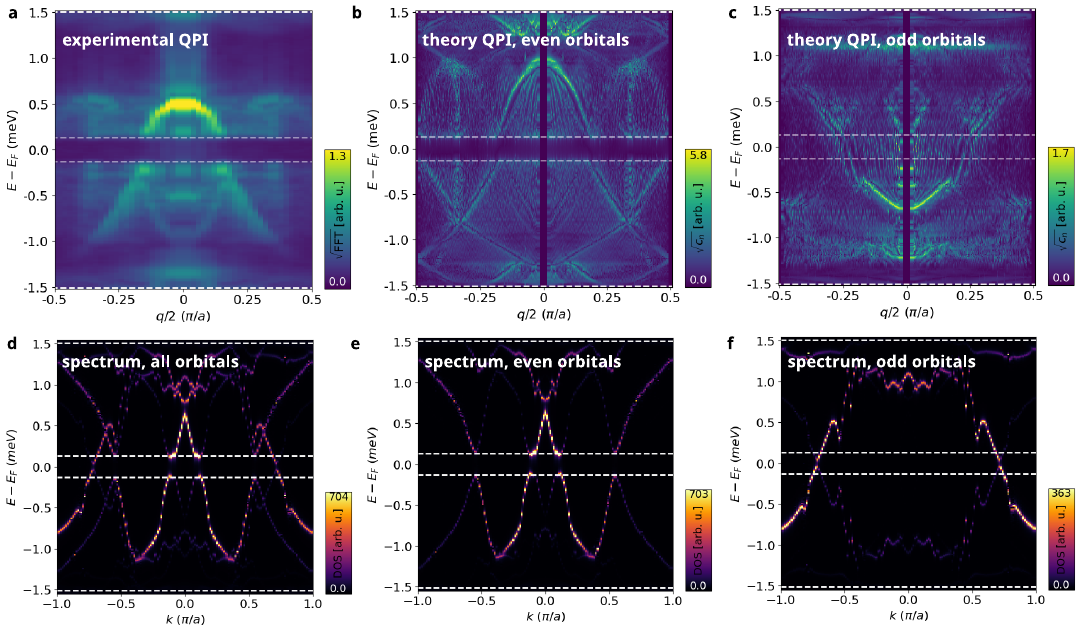}
    \caption{\label{fig:nb-spectrum}
    Spectral properties of Mn chains on Nb(110) from experiments and calculations. \textbf{a}, Experimental QPI dispersion %of scattering vectors 
    extracted from averaging the 1D-FFTs of the deconvoluted \didusig line profiles of Ref.~\cite{Schneider2021a} measured along the center 
    of the $\mathrm{Mn}_{L}$ chains on Nb(110) with $14\leq L \leq36$. %; data from Ref.~\cite{Schneider2021a}. 
    Averaged 1D-FFT of the LDOS of Mn$_{L}$ chains on Nb(110) from \textit{ab initio} calculations with $10\leq L \leq36$ projected to \textbf{b}, the even and \textbf{c}, the odd orbitals. Spectral function of the infinite Mn chain on Nb(110) \textbf{d}, including all orbitals and projected to \textbf{e}, the even and \textbf{f}, the odd orbitals. The white dashed line in each panel at $\pm\Delta_{\textrm{Nb}}=\pm 1.51$~meV indicates the superconducting gap, and at $\pm\Delta_{\textrm{mini,calc}}=\pm 0.13$ meV the minigap observed in the simulations. 
    }
\end{figure*}

The Fourier transforms of the \didusig line profiles obtained along the centers of the chain in Ref.~\cite{Schneider2021a} averaged over the chain lengths are shown in Fig.~\ref{fig:nb-spectrum}a. Since these measurements are mainly sensitive to the even states, they can most directly be compared to the Fourier transforms of the calculated LDOS averaged over the chain lengths and projected on the even states in Fig.~\ref{fig:nb-spectrum}b. The most pronounced feature in the QPI spectra is the parabolic branch with negative curvature at low $q$ values starting at positive energies ($E-E_{\textrm{F}}=0.50$~meV in the experiments in Fig.~\ref{fig:nb-spectrum}a, $E-E_{\textrm{F}}=1.00~\mathrm{meV}$ in the simulations in Fig.~\ref{fig:nb-spectrum}b), which becomes discontinuous at $q/2=\pm 0.17\frac{\pi}{a}$, then can be followed below the minigap at negative energies. Note that less intense features are also observable at higher scattering wave vectors. %in the simulations in Fig.~\ref{fig:nb-spectrum}b in the same energy range between $E-E_{\textrm{F}}=0.13~\mathrm{meV}$ and $E-E_{\textrm{F}}=1.00~\mathrm{meV}$ as the parabolic branch.
%Three branches appear to start 
A trifurcation appears around $q/2=0.34\frac{\pi}{a}$, at energies $E-E_{\textrm{F}}=0.60~\mathrm{meV}$ in the experiments in Fig.~\ref{fig:nb-spectrum}a and $E-E_{\textrm{F}}=1.00~\mathrm{meV}$ in the simulations in Fig.~\ref{fig:nb-spectrum}b, and the resulting three branches proceed toward the minigap. The middle one of these branches is almost vertical in energy, apart from vanishing inside the minigap. %the energy range from $E-E_{\textrm{F}}=-0.80~\mathrm{meV}$ to $E-E_{\textrm{F}}=0.80~\mathrm{meV}$. 
Indications for these multiple branches are also observable in the simulated real-space LDOS in Fig.~\ref{fig:nb-dos}b. For example, the LDOS at $E=0.30$~meV appears to be a superposition of a function with 6-atom-long periodicity (5 maxima, corresponding to $q/2=0.17\frac{\pi}{a}$ on the parabolic branch) and another one with 3-atom-long periodicity (10 maxima, mapping to $q/2=0.34\frac{\pi}{a}$ in the almost vertical branch). %The first one maps to a point on the main parabolic band at around $q/2=0.17\frac{\pi}{a}$, while the second one appears in the almost vertical feature at $q/2=0.34\frac{\pi}{a}$.

For comparison, the band structure of the infinite chain, illustrated by the spectral function determined from the first-principles calculations, is shown for all orbitals in Fig.~\ref{fig:nb-spectrum}d and projected on the even orbitals in Fig.~\ref{fig:nb-spectrum}e. The W shape of the spectrum confirms the presence of multiple states at the same energy as discussed above. These multiple states mean that there is no simple one-to-one correspondence between the wave number along the infinite chain $k$ and the scattering vector $q/2$ obtained from the Fourier transform of the LDOS. For example, the pairwise almost parallel branches of the W explain the formation of the almost vertical feature in the QPI. The topological invariant of the chain, which in this case is the winding number protected by the mirror symmetry, may be deduced by analyzing the avoided band crossings above and below the minigap in the $k>0$ half of the Brillouin zone~\cite{kitaevchain}; see Supplementary Note 3 and Supplementary Figure 3 for a model calculation for the considered systems where the winding number is determined. %The observable two band crossings would give a winding number of two if the effective $p$-wave pairing, which can chosen to be real due to the mirror symmetry, would switch sign between the crossings. However, this would indicate that the pairing would completely vanish at some point between the avoided crossings, and at this wave vector the particle-hole ratio would reach 1, and only one peak would be observable as a function of energy. Although their intensity is different, positive- and negative-energy states are observable at all wave vectors, which does not support this assumption. If the effective $p$-wave pairing does not change sign, the two band crossings cancel each other, and the winding number vanishes. 
The pair of avoided band crossings and the absence of low-energy precursors of MZMs in finite chains observed in Fig.~\ref{fig:nb-dos}f is most consistent with a vanishing winding number. %which leads to the absence of low-energy precursors of MZMs in finite chains, as is indeed observed in Fig.~\ref{fig:nb-dos}f. 
Note that based on the observation of the avoided band crossing at $q/2=\pm 0.17\frac{\pi}{a}$ in the Fourier transform of the \didus in Fig.~\ref{fig:nb-spectrum}a, it was argued in Ref.~\cite{Schneider2021a} that the minigap is topologically non-trivial, although the absence of low-energy states was found puzzling. However, reconstructing the band structure from the QPI data may be difficult, as discussed above in the connection between Fig.~\ref{fig:nb-spectrum}b and d, and this complicates counting the number of avoided crossings in the band structure in the full range of wave vectors required for deducing the topological invariant. The additional low-intensity branches at higher wave vectors in Fig.~\ref{fig:nb-spectrum}a are similar to the features in the simulations in Fig.~\ref{fig:nb-spectrum}b, which point towards a vanishing winding number in the even bands in the experiments as well. %Although additional branches as in Fig.~\ref{fig:nb-spectrum}b are not clearly visible in the Fourier-transformed experimental data in Fig.~\ref{fig:nb-spectrum}a, they may be hidden in the data points observed for $\left|q/2\right|>0.2\frac{\pi}{a}$ in the energy range between $E-E_{\textrm{F}}=0.18~\mathrm{meV}$ and $E-E_{\textrm{F}}=0.50~\mathrm{meV}$. %, and would explain the absence of zero-energy states of even symmetry in the calculations.

The experimental data in Fig.~\ref{fig:nb-spectrum}a also displays very faint crossings at around $q/2=\pm 0.10\frac{\pi}{a}$; see Ref.~\cite{Schneider2021a} for highlighted views of this regime. Based on their %dispersion and 
low intensity at the center of the chain, these may be attributed to the odd states crossing the Fermi level as the chain length is varied in Fig.~\ref{fig:nb-dos}c. In the corresponding calculated LDOS projected on the odd orbitals in Fig.~\ref{fig:nb-spectrum}c, states may also be observed at all energies, and the features with the highest intensity cross the Fermi level at $q/2=\pm 0.25\frac{\pi}{a}$. In the spectral function of the infinite chain projected to the odd orbitals in Fig.~\ref{fig:nb-spectrum}f, a single pair of avoided crossings is observable at $k=\pm0.715\frac{\pi}{a}$, which is backfolded in the Fourier transform of the real-space LDOS. Note that the spectrum for the odd orbitals displays a minigap of size $\Delta_{\textrm{mini,calc,odd}}=0.079$~meV, significantly smaller %in size
compared to the even orbitals. The absence of states in the infinite chain inside this energy range supports the interpretation that the low-energy states observed in the finite chain originate from the boundaries, %only appear due to the free boundaries, 
and may localize at the ends for longer chain lengths. Furthermore, the single avoided crossing at $k>0$ implies that these low-energy states are of topologically non-trivial origin, supporting their interpretation as precursors of MZMs. The different positions of the Fermi-level crossings in the Fourier transforms are connected to the different periodicity with chain length between experiment and simulations in Fig.~\ref{fig:nb-dos}c and e. %The Fermi-level crossings occur at different values of $q/2$ in the experiments and in the calculations, but 
However, we observed in the calculations that the positions of these crossings sensitively depend on the vertical distance between the Mn atoms and the substrate, see Supplementary Note 4 and Supplementary Fig.~4.

The SOC in the system is not only required for opening a minigap in the YSR bands, but it also causes a hybridization between the even and odd states. However, the SOC does not break the mirror symmetry protecting the topological classification~\cite{li2014topological}. If the hybridization between the even and odd states is weak, as supported by the very different features observable between Fig.~\ref{fig:nb-dos}e and f, Fig.~\ref{fig:nb-spectrum}b and c, and Fig.~\ref{fig:nb-spectrum}e and f, it is still justified to treat these bands separately. The winding number of the whole system will correspond to the sum of the winding numbers of the two bands, with the calculations predicting a single winding attributable to the odd states in the present system.

\subsection{Mn chains on Ta(110)}

\begin{figure*}[!t]
     \centering
     \includegraphics[width=1\textwidth]{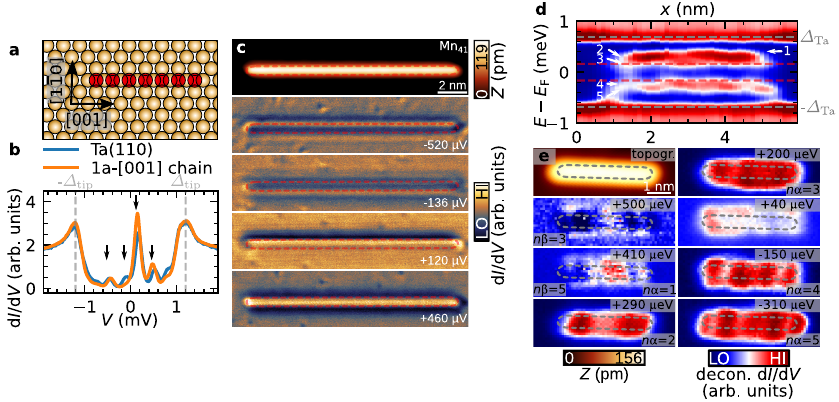}
     \caption{\label{fig:expTa} STM measurements of Mn chains on a Ta(110) substrate. \textbf{a}, Sketch of the arrangement of Ta atoms (yellow) and Mn atoms (red) in the chain. %for a $1a-[001]$ Mn chain on Ta(110). 
     Black arrows indicate the crystallographic directions and are valid for panels \textbf{c} to \textbf{e} as well. \textbf{b}, Comparison of \didus~measured on the Ta(110) substrate and on a $\mathrm{Mn}_{41}$ %$~1a-[001]$ 
     chain using a superconducting Nb tip decorated with Mn atoms. The same microtip was used for panel \textbf{c}, and an out-of-plane magnetic field of $B=+\SI{400}{\milli \tesla}$ was applied. \textbf{c}, Constant-current STM image and constant-contour \didusig maps measured at bias voltages matching the YSR states of the tip (black arrows in panel \textbf{b}). The red dashed lines mark the spatial extent of the Mn chain. The measurement parameters for panels \textbf{b} and \textbf{c} were $V_{\textrm{stab}} = -2$~mV, $I_{\textrm{stab}} = 2$~nA, and $V_{\textrm{mod}} = 40~\mu$V. \textbf{d}, Deconvoluted \didusig line profile measured along the %longitudinal 
     center axis of a $\mathrm{Mn}_{14}$ %1a-[001]$ 
     chain. White arrows and labels indicate the number %\alpha$ 
     of maxima $n$ along the %entire 
     length of the chain. %for a particular energy. 
     Red dashed horizontal lines highlight the edges of the region with reduced intensity. %minigap. 
     \textbf{e} Constant-current STM image and \didusig grid of a $\mathrm{Mn}_{14}$ chain evaluated at energy slices indicated in the top right corner. Gray dashed lines mark the spatial extent of the chain. The measurement parameters for panels \textbf{d} and \textbf{e} were $V_{\textrm{stab}} = -2.5$~mV, $I_{\textrm{stab}} = 1$~nA, and $V_{\textrm{mod}} = 20~\mu$V.
     }
\end{figure*}

Based on our calculations, increasing the strength of the SOC should lead to a localization of the low-energy odd states towards the chain ends, and these states are not expected to be perturbed by the even states due to the formation of the minigap in the same energy range. The increase in the strength of the SOC may be experimentally achieved by replacing the Nb substrate with Ta, a heavier element superconductor with a similar electronic structure. %due to possessing the same number of valence electrons. 
Therefore, we prepared Mn single atoms on the clean (110) surface of a Ta single crystal, and successively built nearest-neighbor Mn$_{L}$ chains along the same $[001]$ direction by STM-tip-induced atom manipulation; see Methods for details. %for the details of sample preparation and measurements, see Ref.~\cite{Beck2023Systematic}.
An illustration of the atomic positions in such a chain is shown in Fig.~\ref{fig:expTa}a. %We picked up Mn atoms with the superconducting Nb tip, leading to the formation of YSR states on the tip. We used this YSR-state-functionalized tip to measure the \didus on the chains. During these measurements, the superconductivity in the substrate was quenched by applying an out-of-plane magnetic field of $B=\SI{400}{\milli \tesla}$, but the tip remained superconducting. This process avoids tunneling between YSR states of the tip and YSR states of the sample~\cite{Huang2020}. Furthermore, it stabilizes the magnetic moment of the tip apex in the field direction. %Thereby, the particle-hole partners of the tip's YSR states have opposite spin orientations~\cite{Cornils2017, Machida2022}. 
The \didus~in Fig.~\ref{fig:expTa}b were measured with Mn atoms attached to the Nb tip on the Ta substrate and a $\mathrm{Mn}_{41}$ chain, in an external magnetic field of $B=\SI{400}{\milli \tesla}$ which quenched superconductivity in the substrate but not in the tip. %indicates the presence of YSR states on the tip. %, and a \didu~measured on the %$\mathrm{Mn}_{41}~1a-[001]$
%$\mathrm{Mn}_{41}$ chain (orange curve) are shown in Fig.~\ref{fig:expTa}(b). 
%The particle-hole partners of the tip's YSR state at $E-E_{\textrm{F}}=\pm \SI{130}{\micro \volt}$ change their intensity asymmetrically when comparing the spectrum on the substrate and on the chain. 
Therefore, the peaks in the spectrum reflect YSR states of the Mn atoms on the tip. The YSR state at $E-E_{\textrm{F}}=\SI{130}{\micro \volt}$ has a higher intensity in the spectrum measured on the $\mathrm{Mn}_{41}$ chain than on the Ta substrate, while the intensity is lower at the negative-bias YSR state with opposite spin polarization~\cite{Cornils2017, Machida2022,Schneider2021c}. This results from magnetoresistive tunneling between the spin-polarized YSR state on the tip and the magnetic chain, enabling to reveal the magnetic structure of the chain with a high signal-to-noise ratio~\cite{Schneider2021c}. %which is a consequence of magnetoresistive tunneling. To determine the spin structure of the chain, 
We measured constant-contour \didusig maps over the %$\mathrm{Mn}_{41}~1a-[001]$
$\mathrm{Mn}_{41}$ chain using this tip, as shown in Fig.~\ref{fig:expTa}c. From the homogeneous increase in signal intensity along the chain at the positive bias voltages matching the YSR states of the tip and from the homogeneous decrease at their negative-bias counterparts, %The \didusig maps reveal an intensity increase (decrease) of the YSR state at $+\SI{130}{\micro \volt}$ ($-\SI{130}{\micro \volt}$) throughout the chain in comparison to the substrate. A similar behavior is observed for the YSR state at $ \pm \SI{500}{\micro \volt}$. Furthermore, we observe that the \didusig signal on the chain is constant in all four \didusig maps. From the increased asymmetry of the tip's YSR states measured on the chain and the absence of any contrast changes along the chain 
we conclude that the chain is in a ferromagnetic state.

We measured the \didusig line profile along the centers of the chains with lengths ranging
%To determine the Shiba band properties we subsequently construct chains with lengths 
from $L=2$ to $L=34$, %and, using a superconducting tip, measure a \didusig line profile for each, 
as shown in Supplementary Movie 3 for the deconvoluted and in Supplementary Movie 4 for the unprocessed raw data. As an example, the deconvoluted \didusig line profile of a %$\mathrm{Mn}_{14}~ 1a-[001]$
$\mathrm{Mn}_{14}$ chain on Ta(110) is shown in Fig.~\ref{fig:expTa}d; see Methods for details. %details of the measurements and of the numerical deconvolution in Ref.~\cite{Beck2023Systematic}. 
Inside the gap of the substrate we find states %dominating in intensity and 
resembling standing waves with increasing numbers of maxima $n$ along the chain at decreasing energies, indicated by white arrows and labels. We identify these as confined YSR states also observed for %For example, we observe $n=1,~2,~3,~4$ and $5$ maxima at energies $+\SI{410}{\micro \electronvolt}$, $+\SI{290}{\micro \electronvolt}$, $+\SI{200}{\micro \electronvolt}$, $-\SI{150}{\micro \electronvolt}$ and $-\SI{310}{\micro \electronvolt}$, respectively. As shown for 
the structurally identical Mn chains on Nb(110) %in experiments~\cite{Schneider2021a} and in the calculations 
in Fig.~\ref{fig:nb-dos}b. %, the standing waves are assigned to confined Bogoliubov quasiparticles in a Shiba band which forms by the hybridization of YSR states of the chain's Mn atoms~\cite{Beck2023Systematic}. 
We find that these states are separated by a region of reduced intensity around the Fermi level, as indicated by the red dashed horizontal lines in Fig.~\ref{fig:expTa}d, which is visible in all \didusig line profiles for chain lengths $N>5$ %(Fig.~\ref{fig:expTa}(d) and 
in Supplementary Movie 3. The high-intensity YSR states are also uncovered by the \didusig grids at the respective energy slices shown in Fig.~\ref{fig:expTa}e. The states at $E-E_{\textrm{F}}=+\SI{410}{\micro \electronvolt}$, $+\SI{290}{\micro \electronvolt}$, $+\SI{200}{\micro \electronvolt}$, $-\SI{150}{\micro \electronvolt}$ and $-\SI{310}{\micro \electronvolt}$ with $n=1,~2,~3,~4$ and $5$ maxima are spatially localized on top of the chain along its center. %the %longitudinal 
%center axis of the chain. 
These states are even under mirroring, and we label them by $n\alpha$, where $\alpha$ refers to the %Therefore, we conclude that the Shiba band producing these confined states primarily stems from hybridization of the 
single-atom YSR state having a maximum on top of the atom~\cite{Beck2023Systematic}. In addition to these %dominant 
states, we observe confined states at $E-E_{\textrm{F}}=+\SI{500}{\micro \electronvolt}$ and $+\SI{40}{\micro \electronvolt}$ resembling odd states because of their %which have an elongated 
intensity minimum directly on top of the atoms along the chain, which are labelled as $n\beta$~\cite{Beck2023Systematic}. %Similarly to Fig.~\ref{fig:nb-spectrum}a, 
These odd states %, c.f. Fig.~\ref{fig:expTa}(e), making them 
are barely visible in the \didusig line profiles along the centers of the chains in Fig.~\ref{fig:expTa}d or Supplementary Movie 3, because their maxima appear %along the \dooe~
on both sides of the chain. %, with an offset in \deeo~that is comparable to the spatial extent of the lobes of the YSR state of the Mn atom denoted by $\beta$ in Ref.~\cite{Beck2023Systematic}, which has a nodal line on top of the atom. 
Note that the distinction between even and odd states is less sharp than on the Nb substrate due to the reduced energy resolution caused by the smaller superconducting gap of Ta, and possibly due to the enhanced hybridization between them because of the stronger SOC. %Therefore, we conclude that these states originate primarily from hybridization of these single atom's YSR states.

\begin{figure*}[!t]
    \centering
    \includegraphics[width=1\textwidth]{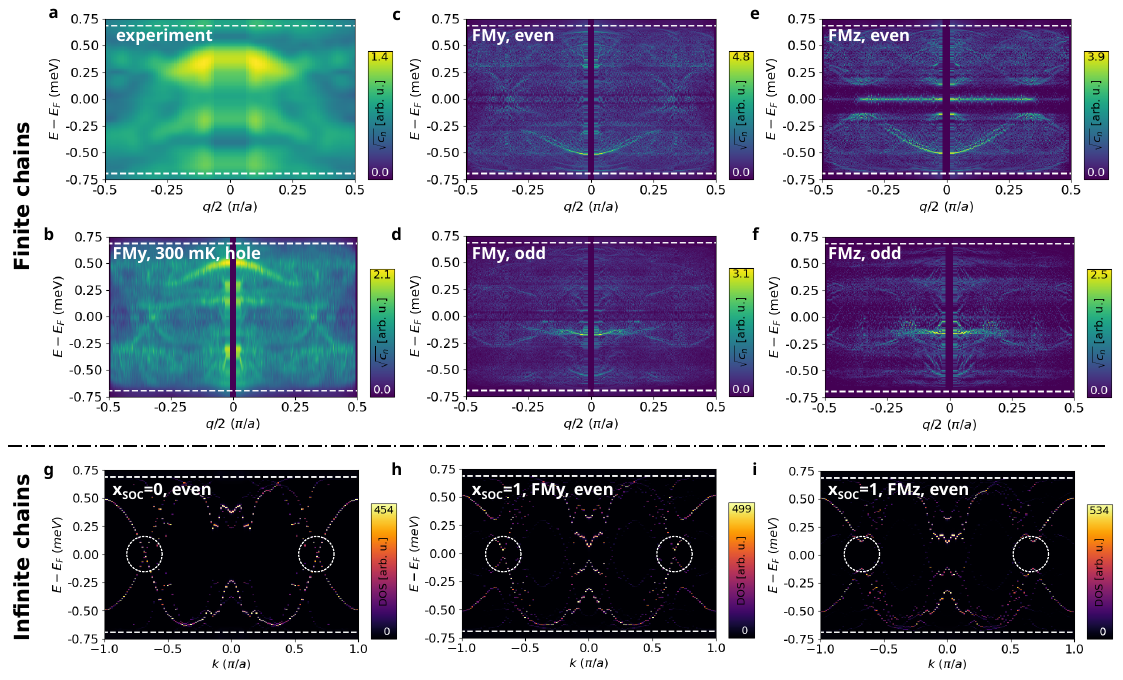}
    \caption{\label{fig:3}
Fourier transform of the LDOS of the Mn chains on Ta(110) from experiments and calculations. \textbf{a}, Averaged 1D-FFT of deconvoluted \didusig line profiles of $\mathrm{Mn}_{L}$ %$~ 1a-[001]$ 
chains on Ta(110) with lengths $14\leq L \leq34$.  %similar to the one shown in Fig.~\ref{fig:expTa}(d).
%Blue and orange dots show %are an overlay of 
%manually evaluated $q(E)$ values from the maxima visible along the center and at the sides of the chain, respectively. 
Averaged 1D-FFT of the LDOS from first-principles calculations: \textbf{b}, hole part, with a Gaussian smearing of $\Delta E/k_{\textrm{B}}=300~\mathrm{mK}$ applied in energy; \textbf{c}-\textbf{f}, electron part of the LDOS projected to even orbitals in panels \textbf{c} and \textbf{e}, and to odd orbitals in panels \textbf{d} and \textbf{f}. The magnetic configuration is ferromagnetic along the chain (FMy) in panels \textbf{b}-\textbf{d} and out-of-plane ferromagnetic (FMz) in panels \textbf{e} and \textbf{f}. %, and with an in-plane ferromagnetic configuration (FMy, along the chain). (c)-(f) 1D-FFT plots with in-plane (FMy) and out-of-plane (FMz) ferromagnetic configurations, projected to even/odd orbitals.
\textbf{g}-\textbf{i}, Spectral function of the infinite Mn chain on Ta(110) projected to the even orbitals \textbf{g}, in the absence of SOC, \textbf{h}, including SOC for in-plane magnetization (FMy) and \textbf{i}, including SOC for out-of-plane magnetization (FMz).  White dashed lines in each panel at $\pm\Delta_{\textrm{Ta}}=\pm0.70$~meV denote the superconducting gap of the substrate. White circles in panels \textbf{g}-\textbf{i} denote the positions of the avoided band crossings.
}
\end{figure*}

To obtain information on the band structure of YSR states in the chains, we followed the procedure for the Nb substrate by averaging the 1D-FFTs of \didusig line profiles for chain lengths $14\leq L \leq34$, 
as shown in  Fig.~\ref{fig:3}a. %Additionally, Fig.~\ref{fig:3}a features an overlay of blue and orange dots which are points of scattering vectors $q/2$ extracted manually from the confined YSR states %of the YSR bands 
%visible in the data (cf. Fig.~\ref{fig:expTa}e) on top and on the side of the chains, respectively. 
%From the comparison of the Fourier transform and the manually evaluated blue dots, 
%we find that the even band has a parabolic shape 
%with negative curvature ranging from $E-E_{\textrm{F}}=+\SI{0.4}{\milli \electronvolt}$ to $E-E_{\textrm{F}}=\SI{-0.600}{\milli \electronvolt}$. 
The highest intensity is observed between $\SI{0.15}{\milli \electronvolt}$ and $\SI{0.60}{\milli \electronvolt}$, and between $-\SI{0.15}{\milli \electronvolt}$ and $-\SI{0.30}{\milli \electronvolt}$, with the features being broader at negative energies. This to some extent resembles the inverted parabolic feature observed for the Nb substrate in Fig.~\ref{fig:nb-spectrum}a, although with a reduced energy resolution. The intensity %of this band 
is reduced between $\pm\SI{0.15}{\milli \electronvolt}$, but it still remains higher than in the lowest-intensity ranges around $\SI{0.60}{\milli \electronvolt}$ and $-\SI{0.30}{\milli \electronvolt}$. %$\pm\SI{0.5}{\milli \electronvolt}$. 
Therefore, the interpretation of this reduced intensity in the low-energy regime as a minigap in the even states is less straightforward than for the Nb substrate. There are identifiable states inside this energy range, see the state at $\SI{40}{\micro \electronvolt}$ in Fig.~\ref{fig:expTa}e; however, these do not resemble end states.

%We also performed 
In the first-principles calculations, %for the Mn$_{L}$ chains on the Ta(110) substrate%, see \hl{Supplementary Note XX} for details
we found a ferromagnetic ground state for the Mn chains on Ta(110) in agreement with the experiments; see Supplementary Note 1. The magnetization was found to lie along the axis of the chain $[001]$ at no external field, in contrast to the out-of-plane magnetized chain on the Nb substrate. We performed the calculations for chain lengths ranging from $L=10$ to $L=37$ atoms (see Supplementary Movie 5 for the real-space data), then took a spatial Fourier transform of the LDOS measured above the chain in the vacuum and averaged over the chain lengths, similarly to the case of the Nb substrate. In Fig.~\ref{fig:3}b, we applied a Gaussian smearing in energy with a width of $\Delta E/k_{\textrm{B}}=300~\mathrm{mK}$, corresponding to the experimental temperature. %It is known that YSR states formed around an impurity show an oscillatory decay with distance inside the substrate with a phase shift between the positive- and negative-energy solutions, meaning that either the positive- or the negative-energy solution may have a higher intensity depending on the measurement position. Our first-principles calculations indicate that such an oscillatory decay and the alternation in intensity can also be observed when moving away from the chain towards the vacuum; see \hl{Supplementary Note 4 and Supplementary Fig.~4}. The intensity ratio between positive and negative energy may also change in the experiments with the distance from the substrate. 
In the calculations we found that the intensity of the features in the LDOS alternates between the positive- and negative-energy parts with increasing vertical distance from the chain; see Supplementary Note 5 and Supplementary Fig.~5. In order to ease the visual comparison with the experimental data, we show the hole part of the calculated LDOS in Fig.~\ref{fig:3}b instead of the electron part, which switches the intensity between positive and negative energies. The most intense feature in the spectrum is a line with negative curvature starting around $E-E_{\textrm{F}}=0.50$~meV at low values of $q/2$, quite close to the experiments where the %maximum of the high-intensity dispersion 
intensity maximum is at around $E-E_{\textrm{F}}=0.40$~meV. The intensity of this branch decreases at lower energies and becomes indistinguishable from the background at around $E-E_{\textrm{F}}=0.27$~meV. The intensity of the Fourier transform is low for all scattering vectors between energies of $\pm0.07$~meV, apart from an apparent band crossing at around $q/2=0.33\frac{\pi}{a}$. Faint lines at negative energies with wave vectors between $q/2=0.33\frac{\pi}{a}$ and $q/2=0.40\frac{\pi}{a}$ appear to be a continuation of the high-intensity feature above the Fermi level. The reduced intensity in the vicinity of the Fermi level with the high-intensity features at positive and negative energies seemingly connected to each other resemble the experimental observations. The states in the vicinity of the Fermi level are not localized towards the chain ends, as shown in Supplementary Movie 5.

The Fourier transform of the LDOS projected on the even and odd states is shown in Fig.~\ref{fig:3}c and d, respectively. We did not include the Gaussian smearing in these figures, and show the electron part of the LDOS. %The maximum intensity of the even and odd orbitals only differs by about 50\%, compared to a factor of almost 4 for the Nb substrate; cf. the maximum values of the color bars. This indicates a stronger hybridization between the even and odd states caused by the stronger SOC, but the features with the highest intensity remain distinguishable. 
The even and odd states hybridize stronger than in the Nb substrate, but the features with the highest intensity remain distinguishable. The parabolic branch identified in Fig.~\ref{fig:3}b can be observed in the even states, although inverted in energy: it starts around $E-E_{\textrm{F}}=-0.50~\mathrm{meV}$ at low $q/2$ values, and its intensity mostly vanishes in the background at $E=-0.27~\mathrm{meV}$ and $q/2=0.28\frac{\pi}{a}$. %Another band is visible 
In the odd orbitals, a broad flat feature around $E-E_{\textrm{F}}=-0.18~\mathrm{meV}$ at low scattering vectors has the highest intensity. %, which is flat for a wide range of $q/2$ values. 
Although the overall intensity is reduced in the vicinity of the Fermi level as discussed above, the superconducting gap is completely filled with states.

To deduce the topological invariant of the bands, we calculated the spectral function of the infinite chain, which is shown projected on the even states without and with SOC in Fig.~\ref{fig:3}g and h, respectively. The dispersion relation approximately resembles the W shape found in Fig.~\ref{fig:nb-spectrum}e for the Nb substrate, but the central peak of the W is now located below the Fermi level. This reduces the number of avoided band crossings for $k>0$ to one at around $k=0.68\frac{\pi}{a}$, as is best visible in the absence of SOC in Fig.~\ref{fig:3}g and highlighted by dashed circles. Including SOC in Fig.~\ref{fig:3}h, the central peak of the W at $k=0$ moves rather close to the Fermi level, but the comparison with Fig.~\ref{fig:3}g and the fact that the effective $p$-wave pairing cannot open a gap at zero wave vector indicates that this is not an avoided crossing. Although the even band appears to possess a single winding, the minigap opened by the SOC at $k=0.68\frac{\pi}{a}$ is $\Delta_{\textrm{Ta,FMy}}=0.03$~meV, which explains why no localized end states are found, rather the full energy range appears to be filled with states for the available chain lengths and energy resolutions in Fig.~\ref{fig:3}a, b and c. Further avoided crossings with low intensity are also visible in Fig.~\ref{fig:3}h, which is a signature of hybridization with the odd band. %Deducing the topological invariant of the odd band is difficult based on the available data
The simulation data for the odd band is most consistent with an odd winding number, see Supplementary Note 6 and Supplementary Figure 6 for a discussion. However, the mirror symmetry still allows for multiple MZMs in the system, meaning that the finite winding number of the even band may give rise to end states regardless of the topological character of the odd band. 

The simulation data indicate that changing the substrate from Nb to Ta introduced a finite winding number in the even band, but the minigap appears to be reduced despite the enhanced SOC.
%The increased SOC of Ta did not improve the localization of the end states in the Mn chains compared to the Nb substrate. The difference between the magnetic structures of the chains is their in-plane and out-of-plane orientation for Ta and Nb substrates, respectively. To explore the role of this difference, 
We repeated the calculations for the Mn chains on Ta(110) in an out-of-plane ferromagnetic alignment, which is the magnetic ground state of the chains on the Nb substrate. 
The Fourier transform of the LDOS based on chain lengths ranging from $L=10$ to $L=37$ projected on the even and odd orbitals are shown in Fig.~\ref{fig:3}e and f, respectively; see Supplementary Movie 6 for the real-space data. %First consider the even orbitals. In this totally relativistic calculations, with low $q$ values we get a 
For the even orbitals, we observe a similar parabolic branch as for in-plane magnetization starting from $E-E_{\textrm{F}}=-0.50~\mathrm{meV}$ at low scattering vectors %, but now it is not that scattered and goes continuously till it reaches 
and continuing to $E-E_{\textrm{F}}=-0.14~\mathrm{meV}$ %, the bottom of the minigap, 
at $q/2=0.34\frac{\pi}{a}$. The intensity between $\pm 0.14~\mathrm{meV}$ is much lower than for the in-plane magnetization, justifying the interpretation of this energy range as a minigap. The parabolic branch appears to continue at positive energies above the minigap, with its particle-hole partner also visible at corresponding negative energies below the minigap. The only feature inside the minigap is a high-intensity bright line at the Fermi level fading outside $q/2=\pm0.34\frac{\pi}{a}$. In the real-space LDOS, this corresponds to zero-energy end states exponentially decaying towards the interior of the chain with a modulation period of around 3 atoms, see Supplementary Movie~6. For the odd states, the flat branch around $E-E_{\textrm{F}}=-0.18~\mathrm{meV}$ is also preserved by rotating the magnetization direction. In contrast to the even states, no minigap may be identified for the odd states. The spectral function of the infinite chain projected on the even states for out-of-plane magnetization in Fig.~\ref{fig:3}i again indicates a single avoided band crossing, which is consistent with the interpretation of the zero-energy end states as MZMs. Note that the main difference compared to the magnetization lying along the chain direction in Fig.~\ref{fig:3}h is the large increase in the minigap to approximately $\Delta_{\textrm{Ta,FMz}}=0.14~\mathrm{meV}$, although the two configurations possess the same symmetries. This increase explains the robustness of the zero-energy end states despite the presence of further low-energy states inside the minigap attributed to the odd states.

\section{Discussion}
In summary, we explored the formation of YSR bands in Mn chains built along the $[001]$ direction on superconducting Nb(110) and Ta(110) substrates using STM/STS measurements and first-principles calculations. These chains are located in a mirror plane, which theoretically enables the coexistence of multiple MZMs at one chain end. Even and odd states with respect to this mirror symmetry may be separated based on their spatial profiles, and can be treated as different bands. 
%On the Nb substrate, we demonstrated that even and odd states with respect to mirroring on the symmetry plane of the chain can be clearly separated. 
In the even band on the Nb substrate, a minigap is opened without any indication for the formation of end states, which our calculations ascribe to a vanishing winding number in this band. %topologically trivial gap opened by the SOC. 
In the odd states, the lowest-lying state oscillates in energy, and increasing the SOC moves this state towards zero energy while making it localized towards the ends of the chains, in agreement with the expectations for precursors of MZMs, and consistent with the single winding deduced from the band structure. On the Ta substrate, we observed a decreased intensity of the YSR states close to the Fermi level, but no clear minigap or end states. The band structure of the even states resembles that of the chains on the Nb substrate, but in this case it can be characterized by a single winding. The absence of observable end states here may be attributed to the very small minigap despite the enhancement of the SOC. %The hybridization between even and odd states is also stronger due to the increased SOC in the substrate. 
Rotating the magnetization direction from the axis of the chain to the out-of-plane direction increases the size of the minigap in the even states, with well-localized end states despite the presence of odd states in the same energy regime. Separately analyzing the topological properties of different YSR bands may support the identification of multiple MZMs which would be difficult to disentangle based simply on spectroscopic data from the end of the chain. Multiple MZMs would enable moving beyond the parity-based characterization of the state proposed for topological qubits based on a single MZM, and could provide information on the influence of the interaction between MZMs on the ground-state degeneracy~\cite{Tewari2012,Fang2014}. %These results highlight the role of the interplay between multi-band effects, the SOC and the magnetic structure on the formation of zero-energy end states in magnetic atomic chains on superconductors.

\section{Methods}

\subsection{STM and STS measurements}
The experimental data for the Nb surface are reproduced from Refs.~\cite{Schneider2021a} and \cite{Crawford2022} where the methods are discussed in detail. Here, we only discuss the measurements on the Ta substrate for which previously unpublished data are presented. Since Ta is the element located one period below Nb in the periodic table, they have a similar electronic configuration of the valence level leading to almost indistinguishable physical properties. They share the body-centered cubic crystal structure, their lattice constants differ only by 0.3\% and their work functions only by 1.5\%~\cite{Michaelson1977}, they have almost identical Fermi surfaces~\cite{Halloran1970,Mattheiss1970}, and both have an occupied $d_{z^{2}}$-like surface state with similar effective masses and binding energies~\cite{Wortelen2015,Eelbo2016,Thonig2016,Obodesko2019}.

All experiments were performed in a home-built ultra-high vacuum STM setup, operated at a temperature of 320 mK~\cite{Wiebe2004}. Constant-current STM images were obtained by applying a bias voltage $V_{\textrm{bias}}$ to the sample, while the tip-sample distance is controlled by a feedback loop such that a constant current $I$ is achieved. \didus were obtained by a standard lock-in technique using a modulation frequency of $f_{\textrm{mod}} = 4142$~Hz and a modulation amplitude referred to as $V_{\textrm{mod}}$ with a typical value of 20~$\mu$V (rms value) added to $V_{\textrm{bias}}$. Prior to obtaining a \didu, the tip was stabilized at $V_{\textrm{stab}}$ and $I_{\textrm{stab}}$. After an initial settling time, the feedback loop was turned off and the bias was swept through a defined range.

\didusig grids and line profiles were obtained by recording \didus on a predefined spatial grid, which was positioned over the structure of interest. \didusig maps are a slice of the grid evaluated at a given bias voltage. Constant-contour \didusig maps were obtained by repeated scanning of individual lines of STM images. In a first sweep each line is measured as it would be the case in a regular constant-current STM image. The $z$-signal of this sweep is saved. In the next sweep, the bias voltage $V_{\textrm{bias}}$ is set to a specific value, the previously recorded $z$-signal is retraced, while the actual feedback is turned off. This allows the measurement of \didusig maps at biases located in the superconducting gap of the sample, which would not be possible using conventional STM images.

The \didus were recorded using a superconducting Nb tip for improved energy resolution. The spectra were deconvoluted to remove the influence of the tip gap and obtain spectra resembling the LDOS of the system, using the procedure described in Ref.~\cite{Beck2023Systematic}. To obtain information about the magnetic ordering of the chain, we picked up Mn atoms with the superconducting Nb tip, leading to the formation of YSR states on the tip, and used this YSR-state-functionalized tip to measure the \didus on the chains~\cite{Schneider2021c}. During these measurements, the superconductivity in the substrate was quenched by applying an out-of-plane magnetic field of $B=\SI{400}{\milli \tesla}$, but the tip remained superconducting. This process avoids tunneling between YSR states of the tip and YSR states of the sample~\cite{Huang2020}. Furthermore, it stabilizes the magnetic moment of the tip apex in the field direction, which leads to opposite spin orientations of the particle-hole partners of the tip’s YSR states~\cite{Machida2022,Schneider2021c}.

\subsection{Sample preparation}
The Ta(110) single crystal was introduced into the ultra-high vacuum chamber and subsequently cleaned by consecutive 30~s long flashes using an e-beam heater at a flashing power of 380~W; see the detailed description of the sample preparation in Ref.~\cite{Beck2023Systematic}. Mn atoms were evaporated to the sample while maintaining a sample temperature below 6~K, to achieve statistically distributed single adatoms. The Mn atoms were reliably positioned by lateral STM-tip-induced atom manipulation~\cite{Eigler1990} at typical tunneling resistances of $\sim 30$~k$\Omega$, depending on the specific microtip. Nanostructures composed of atomically precisely positioned Mn atoms were constructed based on a manipulation image~\cite{Stroscio2004} obtained from moving a single Mn atom over the surface~\cite{Beck2023Systematic}.

\subsection{First-principles calculations}
The first-principles calculations were performed using the fully relativistic screened Korringa--Kohn--Rostoker Green's function code~\cite{Szunyogh1995,csire2018relativistic} in the local spin-density approximation with the Vosko--Wilk--Nusair exchange-correlation potential and the atomic-sphere approximation with an angular-momentum cutoff of $l_{\textrm{max}}=2$. The surface was described by 8 (for Nb) or 7 (for Ta) atomic layers of the substrate and 4 (for Nb) or 5 (for Ta) layers of empty spheres (vacuum) between a semi-infinite bulk substrate and semi-infinite vacuum. The Mn chains were simulated by embedding the row of magnetic atoms along the $[001]$ direction and their substrate and vacuum environment up to %14 
next-nearest-neighbors in the bcc structure in the surface. To obtain an accurate representation of the chains up to a length of 38 atoms, we calculated the surface Green's function of the host system in 7564 $\boldsymbol{k}$ points for the embedding. From the Green's function of the embedded system, we calculated the atomically and energy-resolved local density of states; we present the local density of states calculated for the sites above the magnetic atoms in the vacuum layer. All calculations were performed self-consistently in the normal state, then superconductivity was included by adding a pairing potential on the superconducting atoms to solve the Kohn--Sham--Dirac--Bogoliubov--de Gennes equations, in which case the number of k points was increased to 20604. The pairing potential was chosen such that it reproduces the experimentally observed superconducting gaps in the bulk, $\Delta_{\textrm{Nb}}=1.51~\mathrm{meV}$ and $\Delta_{\textrm{Ta}}=0.69~\mathrm{meV}$. The lattice constants used in the calculations were $a_{\mathrm{Nb}}=330.04~\mathrm{pm}$ and $a_{\textrm{Ta}}=330.29~\mathrm{pm}$. The layer containing the magnetic atoms was relaxed by 4\% on the Nb surface (see Supplementary Note 4 and Supplementary Fig. 4 for different relaxation values) and by 13.4\% on the Ta surface toward the substrate compared to the ideal bulk interlayer distance. The relaxation for the Ta substrate was determined by VASP \cite{Kresse1996,Kresse1996a,Hafner2008} calculations optimizing the geometry of a Mn monolayer on a 4-atomic-layers-thick Ta(110) slab with 2 atoms per layer and 28 \AA\, thick vacuum in the supercell, where a $21\times 21\times 1$ Monkhorst-Pack \cite{Monkhorst1976,Pack1977} $k$-point sampling of the Brillouin zone was used.

The spectral functions presented in the paper are obtained from the generalization of the embedded-cluster method~\cite{Nyari2021} to one-dimensional periodicity. The one-dimensional Green's function of the host system is derived from the two-dimensional Green's function of the layered system by integrating over only the $k$ points perpendicular to the chain. In order to do that, we constructed a rectangular mesh for the Brillouin zone integration along the chain direction. Then we applied the same embedding approach for the one-dimensional Green's function for each $k$ point of the one-dimensional Brillouin zone. The calculations included 151 $k$ points in the one-dimensional Brillouin zone and 215 points for the perpendicular $k$ integration. The imaginary part of the energy %contour 
was $10^{-6}$~Ry for the Nb host and $10^{-7}$~Ry for the Ta host, with 301 energy points in the same range as for finite chains. The one-dimensional unit cell of the embedded wire contained 9 atoms, i.e., the Mn atom with its nearest and next-nearest neighbors in the bcc structure.

\section{Data availability}
The authors declare that the data supporting the findings of this study are available within the paper and its supplementary information files.

\section{Code availability}
The analysis and simulation codes that support the findings of the study are available from the corresponding authors upon reasonable request.

%\emph{Acknowledgements.}
\section{Acknowledgements}
We gratefully acknowledge financial support by the National Research, Development, and Innovation Office (NRDI) of Hungary under Project Nos. K131938, FK142601, 142652 and ADVANCED 149745, by the Ministry of Culture and Innovation and the National Research, Development and Innovation Office within the Quantum Information National Laboratory of Hungary (Grant No. 2022-2.1.1-NL-2022-00004), by the Hungarian Academy of Sciences via a J\'{a}nos Bolyai Research Grant (Grant No. BO/00178/23/11), by the Research Fellowship Programme (Grant No. EKÖP-24-4-II-BME-377) of the Ministry of Culture and Innovation of Hungary from the National Fund for Research, Development and Innovation, by the Hungarian Research Network via Grant No. KMP/2024/48, by the Deutsche Forschungsgemeinschaft (DFG, German Research Foundation) via SFB 925 Project No. 170620586, by the DFG via project WI 3097/4-1 (Project No. 543483081), by the DFG via the Cluster of Excellence ``Advanced Imaging of Matter'' (EXC 2056 - Project No. 390715994), and by the European Union via the ERC Advanced Grant ADMIRE (Project No. 786020). We acknowledge the Digital Government Development and Project Management Ltd. for awarding us access to the Komondor HPC facility based in Hungary.

\section{Author contributions}
All authors discussed the data and approved the manuscript.

\section{Competing interests}
The authors declare no competing interests.

%\section{Materials and correspondence}

%\section{Supplementary information}

\bibliography{references.bib}

\end{document}

\title{Supplementary Information for \\
%TITLE}
%\title{
Coexistence of topologically trivial and non-trivial Yu--Shiba--Rusinov bands in magnetic atomic chains on a superconductor}

\author{Bendeg\'uz Ny\'ari}
\affiliation{Department of Theoretical Physics, Budapest University of Technology and Economics, 1111 Budapest, Hungary}
\affiliation{HUN-REN-BME Condensed Matter Research Group, Budapest University of Technology and Economics, 1111 Budapest, Hungary}
\author{Philip Beck}
%\affiliation{Department of Physics, University of Hamburg, Hamburg, Germany}
\affiliation{Institute of Nanostructure and Solid State Physics, University of Hamburg, 20355 Hamburg, Germany}
\author{Andr\'as L\'aszl\'offy}\email{laszloffy.andras@wigner.hun-ren.hu}
\affiliation{Department of Theoretical Solid State Physics, HUN-REN Wigner Research Centre for Physics, 1525 Budapest, Hungary}
\author{Lucas Schneider}
%\affiliation{Department of Physics, University of Hamburg, Hamburg, Germany}
\affiliation{Department of Physics, University of California, Berkeley, 94720 California, United States}
%\affiliation{Materials Sciences Division, Lawrence Berkeley National Laboratory, Berkeley, 94720 California, United States}
\affiliation{Institute of Nanostructure and Solid State Physics, University of Hamburg, 20355 Hamburg, Germany}
\author{Krisztián Palotás}
\affiliation{Department of Theoretical Solid State Physics, HUN-REN Wigner Research Centre for Physics, 1525 Budapest, Hungary}
%\affiliation{Department of Theoretical Physics, Budapest University of Technology and Economics, 1111 Budapest, Hungary}
%\affiliation{HUN-REN-SZTE Reaction Kinetics and Surface Chemistry Research Group, University of Szeged, 6720 Szeged, Hungary}
\author{László Szunyogh}
\affiliation{Department of Theoretical Physics, Budapest University of Technology and Economics, 1111 Budapest, Hungary}
\affiliation{HUN-REN-BME Condensed Matter Research Group, Budapest University of Technology and Economics, 1111 Budapest, Hungary}
\author{Roland Wiesendanger}
%\affiliation{Department of Physics, University of Hamburg, Hamburg, Germany}
\affiliation{Institute of Nanostructure and Solid State Physics, University of Hamburg, 20355 Hamburg, Germany}
\author{Jens Wiebe}
%\affiliation{Department of Physics, University of Hamburg, Hamburg, Germany}
\affiliation{Institute of Nanostructure and Solid State Physics, University of Hamburg, 20355 Hamburg, Germany}
\author{Bal\'azs \'Ujfalussy}\email{ujfalussy.balazs@wigner.hun-ren.hu}
\affiliation{Department of Theoretical Solid State Physics, HUN-REN Wigner Research Centre for Physics, 1525 Budapest, Hungary}
\author{Levente R\'ozsa}
\affiliation{Department of Theoretical Solid State Physics, HUN-REN Wigner Research Centre for Physics, 1525 Budapest, Hungary}
\affiliation{Department of Theoretical Physics, Budapest University of Technology and Economics, 1111 Budapest, Hungary}

% %% Notice placement of commas and superscripts and use of &
% %% in the author list

% \author{Bendeg\'uz Ny\'ari$^{1,2}$} %https://orcid.org/0000-0001-5524-9995
% \author{Andr\'as L\'aszl\'offy$^{3}$}
% \author{Lucas Schneider}
% \author{Philip Beck}
% \author{L\'aszl\'o Szunyogh$^{1,2}$}
% \author{Jens Wiebe}
% \author{Roland Wiesendanger}
% \author{Bal\'azs \'Ujfalussy$^{3}$}

\maketitle

\section{Spin Hamiltonian and magnetic ground state from first principles}

\begin{figure}
    \centering
    \includegraphics[width=0.8\textwidth]{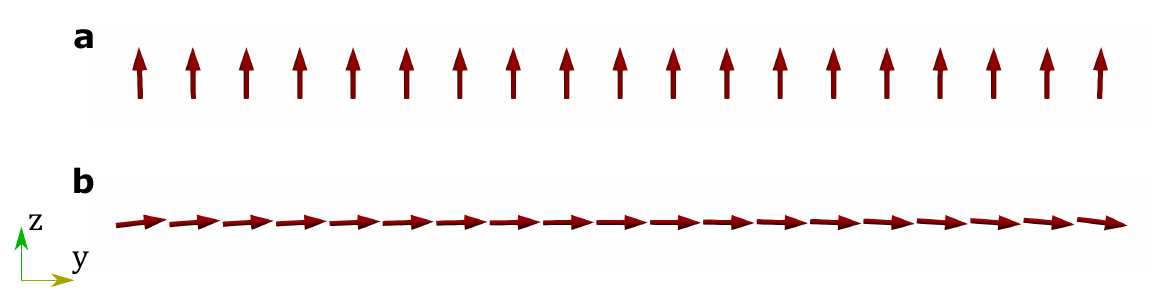}
    \caption{Magnetic ground states of the chains obtained from spin-model simulations. %in the main text. 
    The ground states are shown for a Mn$_{19}$ chain \textbf{a}, on Nb and \textbf{b}, on Ta, respectively. The $y$ axis is along the chain, while the $z$ axis is along the out-of-plane direction. %{\red{In the main text we denoted the chain direction by $x$, it would be easier to change it here than rewriting all the indices there.}}%horizontal axis is along the chain, while the vertical axis is the out-of-plane direction.
    }
    \label{fig:groundstate}
\end{figure}

Relying on the adiabatic decoupling of the electronic and spin degrees of freedom and on the rigid-spin approximation \cite{Antropov1996}, 
the thermodynamic potential of a magnetic system can be parametrized by a set of unit vectors $ \lbrace \vrd{e} \rbrace = \lbrace \vrd{e}_1 , \vrd{e}_2 , \dots , \vrd{e}_N \rbrace $, corresponding to the orientations of the local magnetic moments. %The grand potential $ \Omega \left( \lbrace \vrd{e} \rbrace \right) $ then defines a classical spin Hamiltonian which can be used in numerical simulations. Instead of calculating the grand potential directly, a straightforward idea is to map it 
The thermodynamic potential is mapped onto a generalized Heisenberg model of the form
\begin{equation}
	\Omega\left( \left\{ \vrd{e} \right\} \right) = \Omega_0 + \sum_{i=1}^N \vrd{e}_i \mxrd{K}_i \vrd{e}_i - \frac{1}{2} \sum_{\substack{
   i,j=1 \\
   i \neq j}}^N \vrd{e}_i \mxrd{J}_{ij} \vrd{e}_j,
		\label{eq:heis}
	\end{equation}
where $\Omega_0$ is a constant, $ \mxrd{K}_i $ are second-order single-ion anisotropy matrices, and $\mxrd{J}_{ij}$ are tensorial exchange interactions, which can be decomposed into three parts:
\begin{align}
	\mxrd{J}_{ij} = & J_{ij}^I\mxrd{I} + \mxrd{J}_{ij}^S + \mxrd{J}_{ij}^A ,
\end{align}
	where 
\begin {align}
	J_{ij} =&\frac{1}{3} \Tr \left( \mxrd{J}_{ij} \right) 
\end{align}
	is the isotropic exchange interaction; 
\begin {align}
	\mxrd{J}_{ij}^S = & \frac{1}{2} \left( \mxrd{J}_{ij} + \mxrd{J}_{ij}^T \right) -J_{ij} \mxrd{I},
 \end{align}
with $T$ denoting the transpose of a matrix, is the traceless symmetric part of the matrix which contributes to the magnetic anisotropy of the system (two-ion anisotropy); and the	antisymmetric part of the matrix,
\begin {align}
	 \mxrd{J}_{ij}^A = & \frac{1}{2} \left( \mxrd{J}_{ij}-\mxrd{J}_{ij}^T \right) ,
 \end{align}
is related to the Dzyaloshinskii--Moriya (DM) interaction,
\begin {align}
 \vrd{e}_i \mxrd{J}^A_{ij} \vrd{e}_j = \vrd{D}_{ij} \left(\vrd{e}_i \times \vrd{e}_j \right)
  \end{align}
  with the DM vector $D_{ij}^\alpha = \frac{1}{2} \varepsilon_{\alpha \beta \gamma} J_{ij}^{\beta \gamma}$, $ \varepsilon_{\alpha \beta \gamma}$ being the Levi--Civita symbol. The parameters of the spin model were determined by the spin-cluster expansion in the normal state as implemented in the screened Korringa--Kohn--Rostoker program for finite magnetic structures~\cite{Laszloffy2017}. Here, we discuss the parameter values obtained for neighbors in the middle of the chain, which can be used to understand the magnetic ordering of the chains excluding edge effects.

%The site-resolved effective anisotropy matrix, including single-ion and two-ion contributions, can be defined as \cite{Laszloffy2019}
%\begin{equation}
% \mxrd{A}_{i,\mathrm{FM}/\mathrm{AFM}} = \mxrd{K}_{i} - \frac{1}{2} \sum\limits_{j=1}^{N} \mxrd{J}_{ij}^{S}(\pm 1)^{i+j},
% \label{eq:anisite}
%\end{equation}
%where the sign of $(+1)$ and $(-1)$ has to be used for dimers and chains with (NN) ferromagnetic (FM) and antiferromagnetic (AFM) couplings, respectively.

 %For each system, ten runs with random initial configurations were performed, where we assumed that the actual ground state has been found if at least eight out of the ten runs resulted in the same final state with the lowest energy.

For the Mn$_{19}$ chain on the Nb(110) surface, the nearest-neighbor ferromagnetic isotropic interaction is the strongest, $J^I_{9,10}=20.50~\mathrm{meV}$. The next-nearest-neighbor isotropic interaction $J^I_{9,11}=0.83~\mathrm{meV}$ is considerably weaker, and reinforces the ferromagnetic ordering. The DM vectors only have a finite $x$ component perpendicular to the mirror plane containing the chain. They take the values $D^{x}_{9,10} = -0.43~\mathrm{meV}$ and $D^{x}_{9,11} = 0.14~\mathrm{meV}$ for nearest and next-nearest neighbors, respectively. The total anisotropy energy per spin including both single-ion and two-ion contributions is $\Delta E_{yz}=0.38~\mathrm{meV}$ between the intermediate $y$ and the easy $z$ directions, while it is $\Delta E_{xz}=0.68~\mathrm{meV}$ between the hard $x$ and the easy $z$ axes.

%On the Nb[110] surface, the value of the spin model parameters of the Mn$_{19}$ chain is as follows: the nearest neighbor (NN) isotropic interaction between sites in the middle of the chain is $J^I_{9,10}=20.50~\mathrm{meV}$, the second NN isotropic interaction is $J^I_{9,11}=0.83~\mathrm{meV}$, the amplitude of the NN DM vector reads $D_{9,10} = 0.43~\mathrm{meV}$ and points along $-x$ direction (perpendicular to the chain, as a consequence of the Moriya rules), while for the second NN DM vector it is $D_{9,11} = 0.14~\mathrm{meV}$ and points along $x$ direction, the anisotropy energy (both single-ion and two-ion anisotropy are taken into account) between ferromagnetic configurations in $y$ (medium direction) and $z$ (easy direction) is $\Delta E_{zy}=0.38~\mathrm{meV}$/magnetic site, while the anisotropy energy between the $x$ (hard direction) and the $z$ direction is $\Delta E_{zx}=0.68~\mathrm{meV}$ per magnetic site.

For the Mn$_{19}$ chain on the Ta(110) surface, the nearest-neighbor ferromagnetic isotropic interaction $J^I_{9,10}=42.42~\mathrm{meV}$ is twice as strong as on the Nb surface. The next-nearest-neighbor interaction is antiferromagnetic with $J^I_{9,11}=-0.99~\mathrm{meV}$, but still considerably weaker than the nearest-neighbor term. The $x$ components of the DM vectors are $D^{x}_{9,10} =-3.36 ~\mathrm{meV}$ and $D^{x}_{9,11} =1.35~\mathrm{meV}$ for nearest and next-nearest neighbors, respectively; they are almost ten times stronger than for the Nb substrate due to the enhanced spin-orbit coupling. The easy direction is along the $y$ axis in this case, preferred by $\Delta E_{zy} = 0.28~\mathrm{meV}$ compared to the intermediate $z$ axis. The $x$ axis is energetically even more unfavorable than on the Nb surface, with an energy difference of $\Delta E_{xy} = 3.24~\mathrm{meV}$ per spin compared to the easy axis.

%On the Ta[110] surface and for the Mn$_{19}$ chain the isotropic interactions are $J^I_{9,10}=42.42~\mathrm{meV}$ and $J^I_{9,11}=-0.99~\mathrm{meV}$, the amplitude of the DM vectors are $D_{9,10} =3.36 ~\mathrm{meV}$ and $D_{9,11} =1.35~\mathrm{meV}$ pointing along $-x$ and $x$ directions, respectively, the (ferromagnetic) anisotropy energy between the $z$ (medium direction) and the $y$ (easy direction) directions is $\Delta E_{yz} = 0.28~\mathrm{meV}$/magnetic site, while the anisotropy energy between the $x$ (hard direction) and the $y$ direction is $3.24~\mathrm{meV}$/magnetic site.

The ground state of the magnetic chains is determined from %subsequent 
low-temperature Metropolis Monte Carlo simulations of the spin model, followed by zero-temperature Landau--Lifshitz--Gilbert spin-dynamics simulations started from the final configuration of the Monte Carlo simulations. The details of the %MC 
simulations are given in Refs.~\cite{Laszloffy2017,Laszloffy2019}. 
%The accuracy of the ground state can be improved by the LLG spin dynamics simulations containing the damping term only, 
%and starting from the final state of the MC simulations.\cite{Laszloffy2019}.

In Supplementary Fig.~\ref{fig:groundstate}, we show the ground state of the Mn$_{19}$ chain on both substrates in zero magnetic field. The strong NN isotropic coupling %largely 
results in a ferromagnetic ground state, with the easy axis determined by the anisotropy parameters. The DM interactions between nearest and next-nearest neighbors prefer opposite rotational senses, and %they are weaker than the geometric mean of the isotropic coupling and the anisotropy parameters, meaning that 
they are not strong enough to stabilize a spin-spiral ground state. The effect of the DM interactions can only be observed at the ends of the chain, where the spins are tilted away from the equilibrium direction by $1.6^\circ$ on the Nb surface and by $7^\circ$ on the Ta surface. %, which is an order of magnitude larger than other spin model parameters. The easy anisotropy direction is responsible for that the spins point along the $z$/$y$ direction in the case of the Nb/Ta surface, respectively. The reduced symmetry at the chain ends results in a tilting of the spins compared to the perfect ferromagnetic configuration, especially in the case of the Ta surface, where the magnetic moment of the edge atom is tilted away from $y$ by $7^\circ$ (the same angle is $1.6^\circ$ in the case of the Nb surface).

%%% Old text from the main article:
% The NN isotropic interaction in the middle of the chain is $J_{1}=42$~meV (positive value means ferromagnetic coupling), the second NN interaction is only $J_{2}=-1$~meV. The anisotropy energy between the easy ($y=[001]$, along the chain) and medium ($z=[110]$, perpendicular to the surface) directions is small, $\Delta E_{yz}=0.3$~meV/magnetic site, while the anisotropy energy between the easy and hard ($x=[1\overline{1}0]$, in-plane direction, perpendicular to the chain) directions is $\Delta E_{xy}=3$~meV/magnetic site. %Due to this small value of the magnetic anisotropy energy, the experiments are not able to determine the easy magnetization direction, and we assume that the calculations are correct and the easy direction is the in-plane direction along the chain. 

\section{Spectrum without spin-orbit coupling%Effects of spin-orbit coupling on quasiparticle interference spectra
}
%Turn SOC off, both for Nb and Ta case.
% \begin{figure}[H]
%  	\centering
%      \includegraphics[scale=1]{supp_figs/SFig-SOC_2.pdf}
%      \caption{\label{sfig:SOC_2} Fourier transform of the LDOS on the Mn chains from simulations in the absence of spin-orbit coupling. \textbf{a} and \textbf{b},
%     Averaged 1D-FFT of the LDOS of Mn$_{L}$ chains on Ta(110) from \textit{ab initio} calculations with $14\leq L \leq34$, projected to the \textbf{a}, even and \textbf{b}, odd orbitals. The magnetization points along the [100] or $y$ direction. \textbf{c} and \textbf{d}, Same for Mn$_{L}$ chains on Nb(110) with $10\leq L \leq36$, projected to the \textbf{c}, even and \textbf{d}, odd orbitals. The magnetization points along the [110] or $z$ direction. %\textbf{The effect of SOC on the QPI spectrum}\\ \textbf{a-b} the QPI spetrum of [100] Mn chains on the Ta(110) surface, \textbf{a} without SOC, \textbf{b} with SOC. Both calculated for chains with magnetization along the chain direction. \textbf{c-d} the QPI spetrum of [100] Mn chains on the Nb(110) surface, \textbf{c} without SOC, \textbf{d} with SOC. The Mn atoms have a 4 \% relaxation toward the surface and the chain has an out-of-plane magnetisation.
%      }
%  \end{figure}

\begin{figure}[H]
 	\centering
     \includegraphics[scale=1]{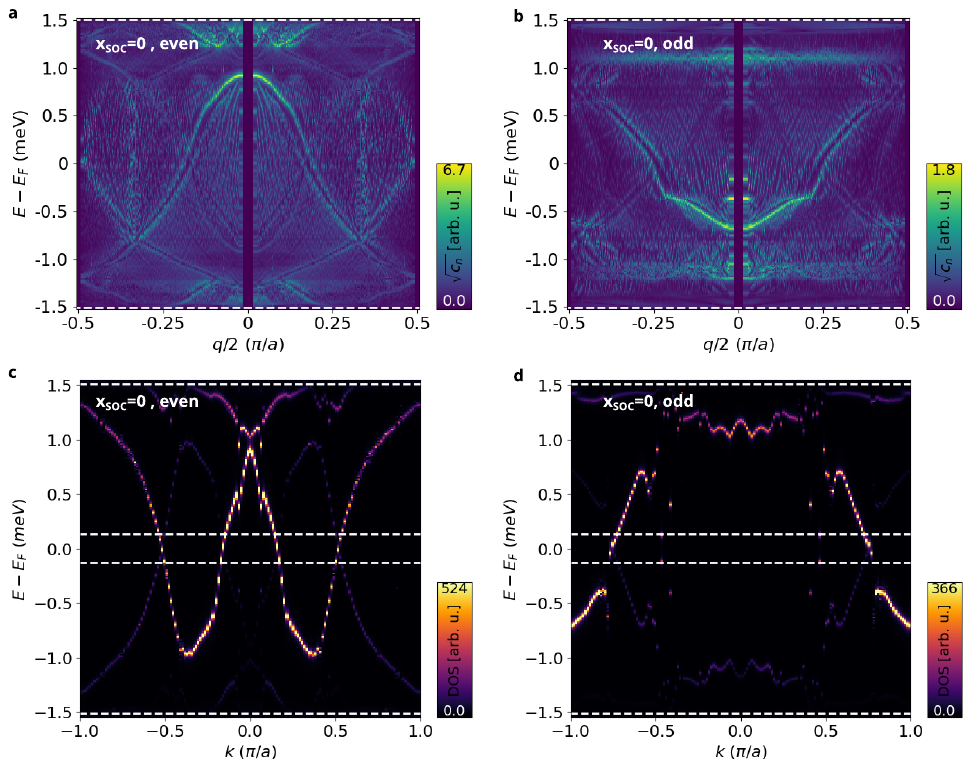}
     \caption{\label{sfig:SOC-Nb} Spectral properties of Mn chains on Nb(110) in the absence of SOC. %Fourier transform of the LDOS on the Mn chains from simulations in the absence of spin-orbit coupling. 
     \textbf{a} and \textbf{b},
    Averaged 1D-FFT of the LDOS of Mn$_{L}$ chains %on Ta(110) from \textit{ab initio} calculations with $14\leq L \leq34$, projected to the \textbf{a}, even and \textbf{b}, odd orbitals. The magnetization points along the [100] or $y$ direction. \textbf{c} and \textbf{d}, Same for Mn$_{L}$ chains 
    on Nb(110) with $10\leq L \leq36$, projected to the \textbf{a}, even and \textbf{b}, odd orbitals. \textbf{c} and \textbf{d}, Spectral function of the infinite chain, projected to the \textbf{c}, even and \textbf{d}, odd orbitals. The magnetization points along the [110] or $z$ direction. %\textbf{The effect of SOC on the QPI spectrum}\\ \textbf{a-b} the QPI spetrum of [100] Mn chains on the Ta(110) surface, \textbf{a} without SOC, \textbf{b} with SOC. Both calculated for chains with magnetization along the chain direction. \textbf{c-d} the QPI spetrum of [100] Mn chains on the Nb(110) surface, \textbf{c} without SOC, \textbf{d} with SOC. The Mn atoms have a 4 \% relaxation toward the surface and the chain has an out-of-plane magnetisation.
     }
\end{figure}

Supplementary Figure~\ref{sfig:SOC-Nb} shows calculation results for the Mn chains on the Nb substrate. The calculations were performed without SOC, but with the same self-consistent potentials and fields as in the relativistic calculations shown in Figs.~1 and 2 in the main text. The results confirm that the minigap in the even states is opened by the SOC, since in its absence the parabolic feature in Supplementary Fig.~\ref{sfig:SOC-Nb}a and the W-shaped band in Supplementary Fig.~\ref{sfig:SOC-Nb}c simply cross the Fermi energy. The effect of turning off the SOC is hardly visible for the odd states in the finite chain %in the case of the Nb substrate 
in Supplementary Fig.~\ref{sfig:SOC-Nb}b compared to Fig.~2c in the main text, while the small minigap in the infinite chain in Fig.~2f in the main text is closed in Supplementary Fig.~\ref{sfig:SOC-Nb}d. %for both the Ta and Nb hosts, see Supplementary Fig.~\ref{sfig:SOC-Nb}a and c, respectively.

\section{Model calculations for the YSR bands}

We use a tight-binding model to simulate the spectral functions and Fourier-transformed LDOS images obtained from first-principles calculations in the main text, which also enables the calculation of the topological invariant. The Hamiltonian reads
\begin{align}
\mathcal{H}=\mathcal{H}_{\textrm{e}}+\mathcal{H}_{\textrm{o}}+\mathcal{H}_{\textrm{hyb}},\label{eq:totalHam}
\end{align}
containing the even $\mathcal{H}_{\textrm{e}}$ and odd $\mathcal{H}_{\textrm{o}}$ bands, and a hybridization $\mathcal{H}_{\textrm{hyb}}$ between them. A single band is described by
\begin{align}
\mathcal{H}_{\textrm{p}}=E_{\textrm{p},0}\sum_{j=1}^{L}c^{\dag}_{\textrm{p},j}c_{\textrm{p},j}+t_{\textrm{p},1}\sum_{j=1}^{L-1}\left(c^{\dag}_{\textrm{p},j}c_{\textrm{p},j+1}+\textrm{h. c.}\right)+t_{2}\sum_{j=1}^{L-2}\left(c^{\dag}_{\textrm{p},j}c_{\textrm{p},j+2}+\textrm{h. c.}\right)+\Delta_{\textrm{p}}\sum_{j=1}^{L-1}\left(c^{\dag}_{\textrm{p},j}c^{\dag}_{\textrm{p},j+1}+\textrm{h. c.}\right),\label{eq:model}
\end{align}
where the parity $\textrm{p}$ denotes either even ($\textrm{e}$) or odd ($\textrm{o}$) orbitals. The $c_{\textrm{p},j}$ spinless fermion operators annihilate single-atom YSR states at site $j$, $E_{\textrm{p},0}$ is the on-site energy, $t_{\textrm{p},1}$ and $t_{\textrm{p},2}$ are hopping parameters, and $\Delta_{\textrm{p}}$ is the $p$-wave pairing. The Hamiltonian for a single band is similar to the model of a $p$-wave superconductor based on which MZMs in wires were first introduced~\cite{kitaevchain}, and has also been successfully applied to describing the YSR bands of Mn chains on Nb(110) in Refs.~\cite{Schneider2021a,Schneider2023}. In the context of YSR bands of ferromagnetic chains, the pairing is only introduced by the SOC, while the hopping parameters can be finite between any pairs of sites even without SOC. We only considered nearest-neighbor and next-nearest-neighbor hopping terms, since these are sufficient to qualitatively reproduce the main features for all bands investigated in the first-principles calculations.

The hybridization term is given by
\begin{align}
\mathcal{H}_{\textrm{hyb}}=\Delta_{\textrm{e--o}}\sum_{j=1}^{L}\left(\textrm{i}c^{\dag}_{\textrm{e},j}c^{\dag}_{\textrm{o},j}+\textrm{h. c.}\right),\label{eq:hyb}
\end{align}
which is also described by a pairing term $\Delta_{\textrm{e--o}}$, since the hybridization between even and odd states is only possible if SOC is taken into account. A similar interband term was considered in Ref.~\cite{Tewari2012}, although for spinful fermions.

Upon expressing the Hamiltonian $\mathcal{H}$ in the Nambu basis, the particle-hole constraint is represented in the usual form as $\mathcal{C}=\tau^{x}\mathcal{K}$, where the Pauli matrix $\tau^{x}$ exchanges the particle and hole subspaces and $\mathcal{K}$ denotes complex conjugation. Note that the parameters $E_{\textrm{p},0}, t_{\textrm{p},1/2},\Delta_{\textrm{p}}$, and $\Delta_{\textrm{e--o}}$ are all required to be real valued. The mirroring on the $yz$ plane is represented as $M=\varrho^{z}$, where $\varrho^{z}$ is a Pauli matrix in even-odd space, expressing that creation and annihilation operators of even states stay invariant under mirroring while for odd states they obtain a negative sign. Together with time-reversal symmetry, represented as $\mathcal{T}=\mathcal{K}$ for spinless fermions, the system possesses an effective time-reversal symmetry $\mathcal{T}_{\textrm{eff}}=M\mathcal{T}$. This symmetry is present because the atoms in the chain are located in a mirror plane, and the magnetic moments also lie in the mirror plane, perpendicular to the surface or along the chain in the considered cases. Although the spin configuration of the chain only indirectly enters the Hamiltonian $\mathcal{H}$, the effective time-reversal symmetry is conserved by all terms, including the hybridization. The combination of the particle-hole constraint and the time-reversal symmetry results in a chiral symmetry $\mathcal{S}=\mathcal{C}\mathcal{T}_{\textrm{eff}}=\tau^{x}\varrho^{z}$.

Due to the chiral symmetry, the system resides in the symmetry class BDI, characterized by the winding number as a topological invariant~\cite{Tewari2012}. In Fourier space, the Hamiltonian may be expressed as
\begin{align}
\mathcal{H}_{k}=\mathcal{H}_{\textrm{e},k}\mathbb{I}_{\textrm{e}}+\mathcal{H}_{\textrm{o},k}\mathbb{I}_{\textrm{o}}-\Delta_{\textrm{e--o}}\tau^{x}\varrho^{y},\label{eq:totalHamk}
\end{align}
where $\mathbb{I}_{\textrm{p}},\textrm{p}\in\left\{\textrm{e},\textrm{o}\right\}$ denotes projection on the even or odd subspace, $\varrho^{y}$ is the corresponding Pauli matrix in even-odd space, and the single-band Hamiltonian is
\begin{align}
\mathcal{H}_{\textrm{p},k}=\left[E_{\textrm{p},0}+2t_{\textrm{p},1}\cos\left(ka\right)+2t_{\textrm{p},2}\cos\left(2ka\right)\right]\tau^{z}-2\Delta_{\textrm{p}}\sin\left(ka\right)\tau^{y},
\end{align}
where $a$ is the atomic spacing of the chain. For a single band, the coefficients of the matrices $\tau^{z}$ and $\tau^{y}$ define a closed curve in two dimensions as $k$ changes from $-\frac{\pi}{a}$ to $\frac{\pi}{a}$, and the topological invariant is the number of times this curve winds around the origin. Here we only discuss nearest-neighbor pairing terms, which restricts the possible values of the winding number to $-1,0$ or $1$. After performing a unitary transformation to a basis where the chiral symmetry is represented as $\mathcal{S}=\tau^{z}$, the total Hamiltonian is rewritten as
\begin{align}
\mathcal{H}_{k}=\left[\begin{array}{cc}0 & A_{k} \\ A^{\dag}_{k} & 0\end{array}\right],
\end{align}
where $A_{k}$ is a $2\times 2$ matrix in even-odd space. For the complete system, the topological invariant is the winding number of the complex phase of $\det A_{k}$ for $k\in\left[-\frac{\pi}{a},\frac{\pi}{a}\right[$~\cite{Tewari2012}. Without hybridization, this corresponds to the sum of the winding numbers of the two bands. The hybridization term $\Delta_{\textrm{e--o}}$ may change the winding number of the complete system, but only by closing and reopening the gap.

% We use a tight-binding model to interpret the Fourier transforms of the LDOS obtained for the even states in Fig.~2b in the main text. As it was shown in Ref.~\cite{Schneider2021a,Schneider2023}, the spectrum of a single YSR band close to the Fermi level may be well approximated by a tight-binding model,
% \begin{align}
% \mathcal{H}=E_{0}\sum_{j=1}^{L}c^{\dag}_{j}c_{j}+t_{1}\sum_{j=1}^{L-1}\left(c^{\dag}_{j}c_{j+1}+\textrm{h. c.}\right)+t_{2}\sum_{j=1}^{L-2}\left(c^{\dag}_{j}c_{j+2}+\textrm{h. c.}\right)+\Delta_{p}\sum_{j=1}^{L-1}\left(c^{\dag}_{j}c^{\dag}_{j+1}+\textrm{h. c.}\right),\label{eq:model}
% \end{align}
% where the $c_{i}$ spinless fermion operators annihilate single-atom YSR states, $E_{0}$ is the on-site energy, $t_{1}$ and $t_{2}$ are hopping parameters, and $\Delta_{p}$ is the $p$-wave pairing. In the ferromagnetic configuration considered here, the hopping parameters are finite between any two sites without SOC, while the pairing is only induced by the SOC. We only considered nearest-neighbor and next-nearest-neighbor hopping terms, since these are sufficient to qualitatively reproduce the main features observed in the calculations.

The spectrum of the infinite system obtained from the eigenvalues of $\mathcal{H}_{k}$ is compared to the LDOS calculated for the finite chain. The formula for the LDOS is
%The LDOS is calculated as
\begin{align}
\textrm{LDOS}_{\textrm{p}}\left(E,j\right)=-\frac{1}{\pi}\lim_{\delta\rightarrow 0+}\textrm{Im}\textrm{Tr}\left(G_{jj}\left(E+\textrm{i}\delta\right)\frac{I_{4}+\tau^{z}}{2}\mathbb{I}_{\textrm{p}}\right),
\end{align}
where $G\left(z\right)=\left(z-H\right)^{-1}$ is the Green's function, $H$ is the matrix of the Hamiltonian $\mathcal{H}$, $G_{jj}$ is the site-diagonal block of the Green's function, which is then projected to the electron part and the band with the selected parity. The LDOS is calculated the same way as in the SKKR method, but the Green's function is different between the two cases. %The Fourier transformation is also performed using the same procedure. 
% For comparison, we also determine the band structure of the chain with periodic boundary conditions, which simply reads
% \begin{align}
% E_{\pm}\left(k\right)=\pm\sqrt{\left(E_{0}+2t_{1}\cos k+2t_{2}\cos 2k\right)^{2}+\left(2\Delta_{p}\sin k\right)^{2}},
% \end{align}
% with $k\in\left[-\pi,\pi\right[$.

\begin{figure}
 	\begin{center}
     \includegraphics[width=\columnwidth]{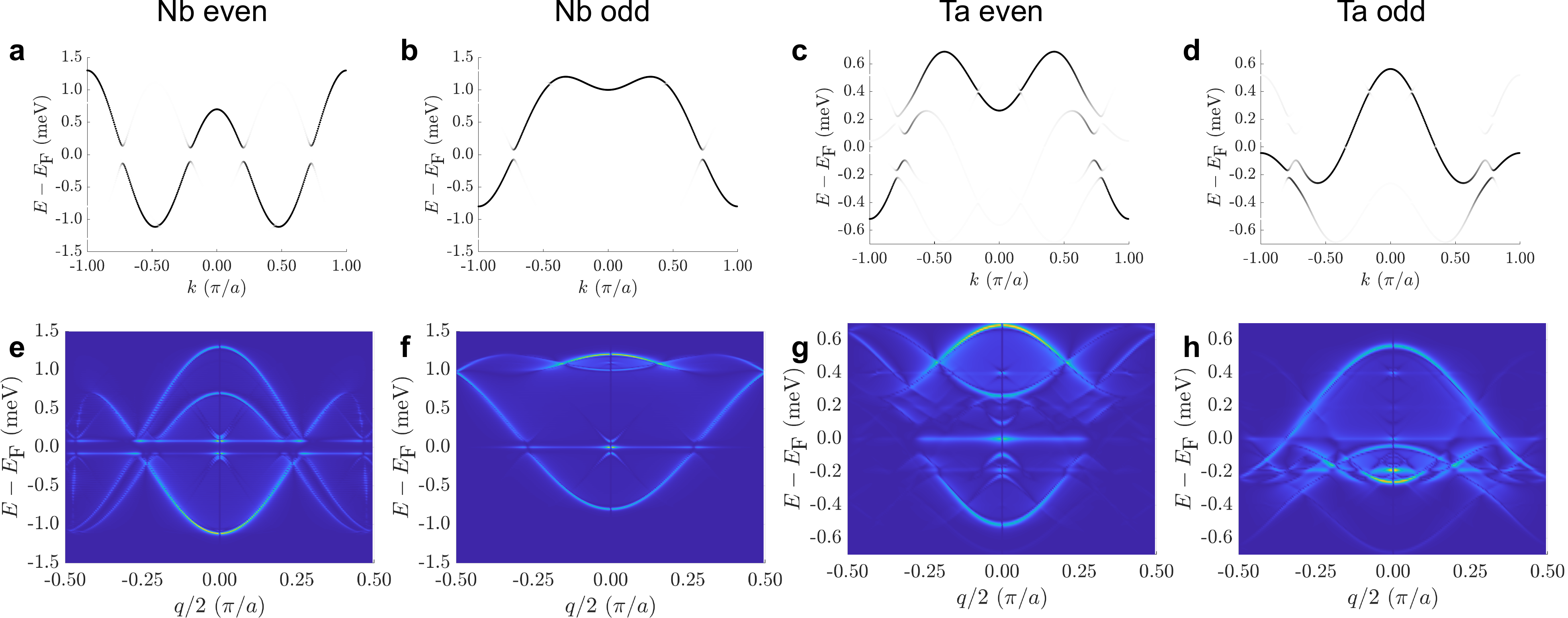}
     \end{center}
     \caption{\label{fig:sfig1} Model calculations for the YSR band structure. \textbf{a}-\textbf{d}, Spectrum of the chain with periodic boundary conditions for model parameters representing the Mn chain on Nb and Ta surfaces. \textbf{e}-\textbf{h}, Fourier transform of the LDOS of a finite chain of length $N=200$. The intensity represents the projection onto the electron part and the even or the odd band as indicated. The parameters in Eq.~\eqref{eq:totalHam} are $E_{\textrm{e},0,\textrm{Nb}}=-0.0446$~meV, $t_{\textrm{e},1,\textrm{Nb}}=-0.1500$~meV, $t_{\textrm{e},2,\textrm{Nb}}=0.5223$~meV, $\Delta_{\textrm{e},\textrm{Nb}}=0.0900$~meV, $E_{\textrm{o},0,\textrm{Nb}}=0.5331$~meV, $t_{\textrm{o},1,\textrm{Nb}}=0.4500$~meV, $t_{\textrm{o},2,\textrm{Nb}}=-0.2166$~meV, $\Delta_{\textrm{o},\textrm{Nb}}=0.0500$~meV, $\Delta_{\textrm{e--o},\textrm{Nb}}=0.0100$~meV; $E_{\textrm{e},0,\textrm{Ta}}=0.2383$~meV, $t_{\textrm{e},1,\textrm{Ta}}=0.1875$~meV, $t_{\textrm{e},2,\textrm{Ta}}=-0.1817$~meV, $\Delta_{\textrm{e},\textrm{Ta}}=0.0800$~meV, $E_{\textrm{o},0,\textrm{Ta}}=0.0010$~meV, $t_{\textrm{o},1,\textrm{Ta}}=0.1438$~meV, $t_{\textrm{o},2,\textrm{Ta}}=0.1307$~meV, $\Delta_{\textrm{o},\textrm{Ta}}=-0.0050$~meV, and $\Delta_{\textrm{e--o},\textrm{Ta}}=0.1000$~meV.%$E_{\textrm{e},0,\textrm{Nb}}=0.0023$~meV, $t_{\textrm{e},1,\textrm{Nb}}=-0.1000$~meV, $t_{\textrm{e},2,\textrm{Nb}}=0.5489$~meV, $\Delta_{\textrm{e},\textrm{Nb}}=0.0900$~meV, $E_{\textrm{o},0,\textrm{Nb}}=0.6756$~meV, $t_{\textrm{o},1,\textrm{Nb}}=0.4375$~meV, $t_{\textrm{o},2,\textrm{Nb}}=-0.2753$~meV, $\Delta_{\textrm{o},\textrm{Nb}}=0.0950$~meV, $\Delta_{\textrm{e--o},\textrm{Nb}}=0.0100$~meV; $E_{\textrm{e},0,\textrm{Ta}}=0.2208$~meV, $t_{\textrm{e},1,\textrm{Ta}}=0.1625$~meV, $t_{\textrm{e},2,\textrm{Ta}}=-0.1979$~meV, $\Delta_{\textrm{e},\textrm{Ta}}=0.0800$~meV, $E_{\textrm{o},0,\textrm{Ta}}=-0.0258$~meV, $t_{\textrm{o},1,\textrm{Ta}}=0.1438$~meV, $t_{\textrm{o},2,\textrm{Ta}}=-0.1442$~meV, $\Delta_{\textrm{o},\textrm{Ta}}=-0.0050$~meV, and $\Delta_{\textrm{e--o},\textrm{Ta}}=0.1000$~meV. %\textbf{a}, Fourier transform of the LDOS of a chain of length $L=100$ and \textbf{b}, spectrum of the same chain with periodic boundary conditions. The parameters in Eq.~\eqref{eq:model} are $E_{0}=-4.0$~meV, $t_{1}=2.5$~meV, $t_{2}=0.0$~meV, $\Delta=0.12$~meV. \textbf{c} and \textbf{d}, Same as in panels \textbf{a} and \textbf{b} using the different parameter set $E_{0}=-0.3$~meV, $t_{1}=0.3$~meV, $t_{2}=-0.6$~meV, $\Delta=0.12$~meV. The regularization parameter of the Green's function was $\delta=0.01$~meV.
     }
\end{figure}

The results of the model calculations are displayed in Supplementary Fig.~\ref{fig:sfig1}. The model parameters were fitted to reproduce the main features of the spectral functions determined from first-principles calculations in the main text. In particular, all spectral functions separately for the even and odd bands may be approximated by a W or M shape, and the energy positions of the maxima and the minima allow for determining the parameters $E_{\textrm{p},0},t_{\textrm{p},1},$ and $t_{\textrm{p},2}$. The pairing parameters $\Delta_{\textrm{p}}$ were chosen to approximately reproduce the size of the minigap. We also considered a term $\Delta_{\textrm{e--o}}$ to illustrate the hybridization of the even and odd bands observable in the first-principles simulations; this slightly affected the gap sizes but did not change the topological invariant, even for Ta where a relatively large value was assumed. For the even band on the Nb substrate in Supplementary Fig. \ref{fig:sfig1}a and e (cf. Fig. 2e and b in the main text), a pair of avoided band crossings may be observed in the dispersion relation, which results in a vanishing winding number. Consequently, a minigap without low-energy states can be clearly identified in the Fourier transform of the LDOS. Note that the additional faint lines observable at higher $q/2$ values %at positive energies 
in Fig.~2b of the main text are also reproduced here, which do not appear without a double avoided crossing; cf.~Supplementary Fig.~\ref{fig:sfig1}f as an example. The pair of avoided crossings could also represent a double winding, which can be included in the model by considering next-nearest-neighbor pairing terms. However, in the model calculations we found that a nonzero winding number always leads to the emergence of low-energy states inside the minigap, even for short chain lengths where they are not necessarily at zero energy, which is incompatible with the results of the first-principles simulations and the experiments. The odd band on the Nb substrate in Supplementary Fig. \ref{fig:sfig1}b and f (cf. Fig. 2f and c in the main text, also Supplementary Fig.~\ref{sfig:SOC-Nb}b and d) displays only a single avoided crossing at high wave vector, leading to a single winding. For the chain length of $L=200$ in the model calculations, this leads to a well-localized end state close to zero energy, showing up as a line at zero energy in the Fourier transform of the LDOS, similarly to the model calculations in Refs.~\cite{Schneider2021a,Crawford2022}. For the shorter chain lengths $L<40$ available in experiments and first-principles simulations, this state moves away from zero energy and changes its energy position with the chain length. 

For the even band on the Ta substrate in Supplementary Fig. \ref{fig:sfig1}c and g (cf. Fig. 4i and e in the main text), again only a single avoided band crossing may be identified in the spectrum, leading to a zero-energy end state visible in the Fourier transform of the LDOS. The pairing here was chosen to reproduce the minigap observed for the out-of-plane magnetization in the main text. The odd band on the Ta substrate in Supplementary Fig. \ref{fig:sfig1}d and h (cf. Supplementary Fig. 6 and Fig. 4f in the main text) similarly shows a single avoided band crossing, but the minigap is very small in this case. Consequently, no clear zero-energy end states can be observed in the Fourier transform of the LDOS even for relatively long chain lengths. Note that particle-hole ratios are close to being equal in the first-principles simulations for the Ta substrate, which is not matched by the model calculations, and the large curvature of the odd band could not be well reproduced by the next-nearest-neighbor model, either. However, the strong hybridization between the bands is qualitatively well captured by the high value of $\Delta_{\textrm{e--o},\textrm{Ta}}$, and the good localization of the end state attributed to the even band in Supplementary Fig. \ref{fig:sfig1}g persists despite the presence of odd states in this energy range. Note that with these parameters, the total Hamiltonian has a double winding. Switching the sign of $\Delta_{\textrm{o},\textrm{Ta}}$ in the model calculations results in opposite winding numbers for the two bands, and a cancellation in the total Hamiltonian; however, in this case the hybridization term moves the lowest-lying state visible in the even bands away from zero energy even for the long chain considered here. This indicates that the observation of a state at nearly zero energy in the first-principles simulations is more consistent with a topologically non-trivial origin resulting from a finite winding number.

\section{The effect of impurity relaxation on the YSR %Shiba
band structure %in case of Mn/Nb(110)
}

The vertical position of an impurity on the surface %can affect many physical properties, such as the charge transfer between the impurity and the surface or the magnetic moment of the impurity. For magnetic impurities on superconductors, it 
can change the position of the YSR peaks inside the superconducting gap by modifying the Kondo coupling to the surface or the magnetic moment of the impurity. %As we set up a calculation, it is always required to obtain the vertical position of the impurity above the surface. %and one also has to take into account the reordering of the surface around the impurity. 
We illustrate this effect on the Mn chains on Nb(110) by changing the relaxation of the layer containing the Mn sites from 0\% to 8\% towards the surface with respect to the bulk Nb interlayer distance in steps of 4\%. We also considered a small relaxation of $-3.6\%$ for the top layer of Nb obtained theoretically from VASP geometry optimization. %(see Methods).

%Within the KKR method we are dealing with layered systems, hence we can only set the position of an atomic layer, not the position of the impurity sites separately. There are many different principles on how to choose the relaxation correctly. In some theoretical methods, optimal structures can be found by calculating the forces acting on the atoms and changing the positions to eliminate forces. In this study, we used an alternative way, fitting the experimental band structure of [100] Mn chains on Nb(110). We are changing the distance between the top Nb layer and the vacuum layer, where the Mn sites are embedded, and we calculate the QPI spectrum in each case. The relaxation of the Mn chains is considered to be correct if we are closest to the experimental spectrum. In our calculations, we changed the relaxation of the Mn sites from 0 \% to 20 \% toward the surface in 4 \% steps, meaning that the distance between the top Nb layer and the Mn chains is decreased with the given amount compared to the ideal bulk position. We also considered a small relaxation of $-3.6 \%$ for the top layer of Nb obtained theoretically from VASP geometry optimization and related to the relaxation of the pure Nb surface. 

The spectral features of the chains are compared in Supplementary Fig.~\ref{sfig:relax} between different relaxation values. The Fourier transforms of the LDOS in Supplementary Fig.~\ref{sfig:relax}a-c display a parabolic feature with a high intensity, which is pushed away from the Fermi level for increasing relaxation. The LDOS is reduced in the vicinity of the Fermi level both for 0\% and 4\% relaxation. A comparison with the calculated spectral functions in Supplementary Fig.~\ref{sfig:relax}d and e reveals that this has a different origin in the two cases: while the even band does not cross the Fermi level for 0\% relaxation, a pair of avoided band crossings can be observed for 4\% relaxation as the bottom of the W-shaped dispersion gets pushed across the Fermi level, as also shown in Fig.~2d in the main text. For 8\% relaxation in Supplementary Fig.~\ref{sfig:relax}f, the avoided crossings get closer to $k=0$ as the middle of the W approaches the Fermi level, and the minigap opened by the SOC reduces, making it no longer identifiable in Supplementary Fig.~\ref{sfig:relax}c. The odd band with minima at the edge of the Brillouin zone does not cross the Fermi level for 0\% relaxation, but is rather close to it. This minimum also gets pushed across the Fermi level as the hybridization of the chain with the substrate is increased at 4\% relaxation, where the avoided crossing is already at $k=0.715\frac{\pi}{a}$. Somewhere between 0\% and 4\% relaxation values this crossing is most likely located close to $k=0.90\frac{\pi}{a}$, which would show up at $q/2=0.10\frac{\pi}{a}$ in the Fourier transforms, close to the position of the low-energy states observed in the experiments in Fig.~2a of the main text. The avoided crossing stemming from the odd bands is found at an even lower wave vector of $k=0.70\frac{\pi}{a}$ for 8\% relaxation. These data demonstrate that the energies of the states, the positions of the avoided crossings and the minigap sizes are very sensitive to the details of the electronic structure, but the shape of the band which is decisive for determining the winding number remains robust.

\begin{figure}
 	\centering
     \includegraphics[scale=1]{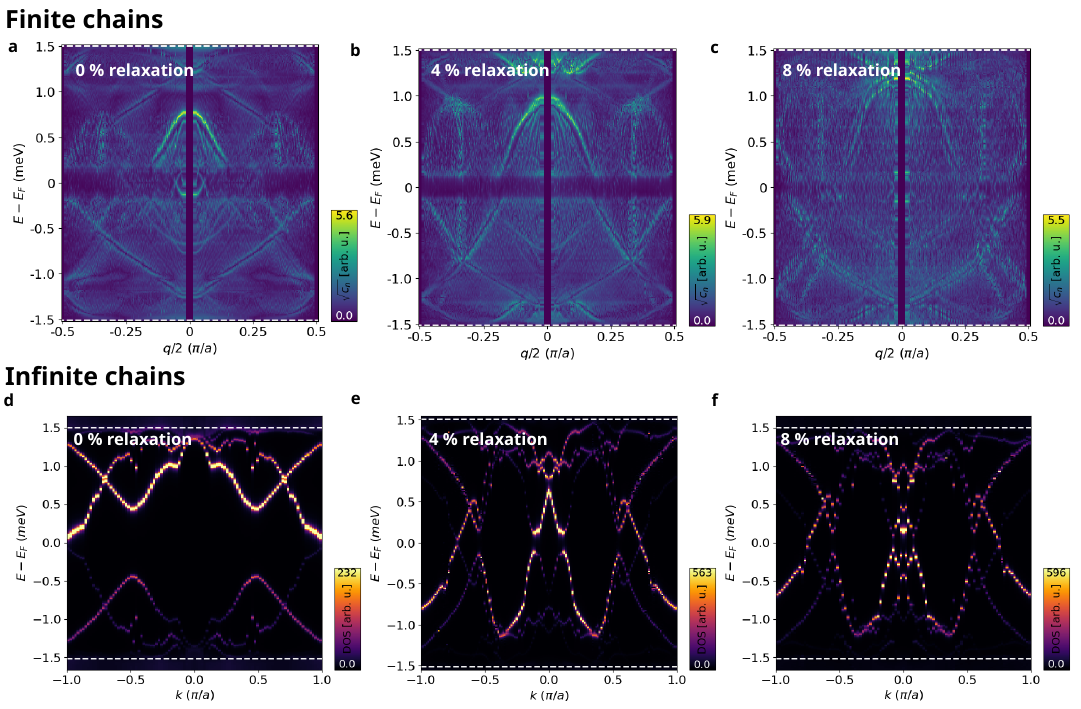}
     \caption{\label{sfig:relax} Spectral properties of Mn chains on Nb(110) for different vertical distances between the magnetic atoms and the substrate. %Fourier transform of the LDOS on the Mn chains from simulations at different spatial positions. 
     \textbf{a-c}, Averaged 1D-FFT of the LDOS of Mn$_{L}$ chains on Nb(110) %from \textit{ab initio} calculations 
     with $10\leq L \leq39$. The relaxation values are indicated in the panels. \textbf{d-f}, Spectral function of the infinite chain for the same relaxation values. %\textbf{Relaxation of [100] Mn chains on the Nb(110) surface }\\ The QPI spectrum of [100] Mn chains on the Nb(110) with the different values of Mn relaxaions. Calculated from chain with length ranging from 10 to 39.
     }
\end{figure}
%\section{Electron-hole oscillation with vertical distance\label{ssec:ehosc}}
\section{Electron-hole oscillations of YSR states with vertical distance\label{ssec:ehosc}}
The intensity of YSR states in the bulk shows an oscillating algebraic decay from the magnetic impurity, with the period of the oscillation determined by the Fermi surface of the superconductor~\cite{yu1965bound,rusinov1969superconductivity}. At longer distances, an exponential decay with the length scale determined by the binding energy of the YSR state and the Fermi velocity is observed. The binding energy also leads to a phase difference in the oscillations, and consequently an asymmetry in intensity, between the positive- and negative-energy solutions, or equivalently the electron and hole parts of the solution at a fixed energy.

%The Yu--Shiba--Rusinov states appear as pairs, one at negative and one at positive energy (relative to the Fermi level) and show asymmetric intensities. In the context of the Bogoliubov--de Gennes equation, it can be described with the electron hole oscillation. The electron and hole intensities of the YSR states vary in space differently, meaning that the electron hole ratio is not a constant and varies as we move away from the impurity. The same argument holds for the Mn chains on Nb and Ta sufaces, as we illustrate by calculating the spectrum from the DOS of the magnetic atoms and from the DOS calculated in different vacuum positions around the chain, see. %
%
%As illustated in
In our simulations, we observe the distance-dependent oscillations also in the vacuum above the surface, accompanied by a fast exponential decay. The LDOS calculated on the Mn atoms and in the first two vacuum layers above the chain are shown in Supplementary Fig.~\ref{sfig:eh-osc}, after performing Fourier transformation and averaging over different chain lengths as in Figs.~2 and 4 of the main text. We used the out-of-plane magnetized and along-the-chain magnetized ground states for Nb and Ta surfaces, respectively. %SFig.~\ref{sfig:eh-osc}. The QPI spectrum changes drastically depending on which atomic positions are used to calculate it. 

%which has a very low contrast in the other figures, but its e-h partner can be clearly seen at positive energy on the Mn sites at $+$0.5 meV. The flat band at $-$0.12 meV is more pronounced on the Mn sites than in the first vacuum layer meaning that it has a large electron component at negative energy, however, in the second vacuum layer there are similar intensities at positive and negative energies in this energy region, resulting in an electron-hole ratio of around 1.

For the Nb host, the most pronounced feature in the LDOS in the first vacuum layer in Supplementary Fig.~\ref{sfig:eh-osc}b is the parabolic feature starting at $E-E_{\textrm{F}}=$1.00 meV, originating from the even states in Fig.~2b in the main text. The electron-hole partner of this feature at negative energies is more intense in the LDOS calculated directly on the Mn atoms in Supplementary Fig.~\ref{sfig:eh-osc}a. In the same panel, two V-shaped features of similar intensity may be observed. The one at negative energy starting at $E-E_{\textrm{F}}=-$0.70 meV comes from the odd states in Fig.~2c in the main text, although it has a relatively lower intensity compared to the even states in the first vacuum layer. The positive-energy V-shaped branch starts at $E-E_{\textrm{F}}=$0.90 meV, and its electron-hole partner is also faintly visible in the first vacuum layer in Supplementary Fig.~\ref{sfig:eh-osc}b where it is not suppressed by the high-intensity parabolic band. This feature may also be identified in Fig.~2b of the main text, meaning that it mainly originates from the even states. On the vacuum sites to the side along the line parallel to the chains in Supplementary Fig.~\ref{sfig:eh-osc}c, the odd states are more pronounced since they have a nodal line in the mirror plane containing the Mn atoms. The Fermi-level crossings of the odd states are much more visible than in Supplementary Fig.~\ref{sfig:eh-osc}a and b, and the minigap is filled. The even states have a smaller extension perpendicular to the chains; consequently, the parabolic feature attributed to the even states has a reduced intensity in Supplementary Fig.~\ref{sfig:eh-osc}c, while the V-shaped branch is relatively more pronounced. Note that the flat features observable above $E-E_{\textrm{F}}=$1.00 meV on the Mn sites obtain a higher intensity for opposite energies below $E-E_{\textrm{F}}=-$1.00 meV on the vacuum atoms next to the chains.

% The case of the Nb host is even more spectacular. On the Mn sites there is a reversed `V'-like band at positive energy, a parabolic band at negative energy and an other `V'-like band at negative energy. Moving to the first vacuum layer the parabolic band is at positive energy and both `V'-shaped band (with much smaller intensity compared to the parabolic band) appear at negative energy, the one starting from lower energy and with larger slope is barely visible. The reversion of the bands make evident that electron-hole oscillation can be observed in this system.

In the case of the Ta host, the Fourier transform from the first vacuum layer in Supplementary Fig.~\ref{sfig:eh-osc}e displays an intensive parabolic feature starting at $E-E_{\textrm{F}}=-$0.50 meV, mainly associated with the even states in Fig.~4c. In the LDOS calculated directly on the Mn atoms in Supplementary Fig.~\ref{sfig:eh-osc}d, the electron-hole partner of the same feature is more pronounced starting at $E-E_{\textrm{F}}=$0.50 meV. On the Mn atoms, the flat feature around $E-E_{\textrm{F}}=-$0.12 meV is the most intense, which comes from the odd states shown in Fig.~4d in the first vacuum layer also at negative energies. In the second vacuum layer in Supplementary Fig.~\ref{sfig:eh-osc}f the intensities at positive and negative energies are similar.% in this energy region.

% At the first look, the spectrum from the Mn sites and from the vacuum positions from the same layer next to the Mn sites (line parallel to the chain, off-side) are very similar. Some minor differences can be observed due to the symmetry of the different bands: the even states are more pronounced off-side the chain having a nodal line in the mirror plane containing the Mn atoms. As a consequence, the Fermi level crossings of the even band is much more visible and fills the minigap. Note that some branches of the spectrum that are above 1 meV on the Mn sites reverse as we move next to the chain, they are more visible below $-1$ meV. 
%A similar effect can be observed if we move off-side of the chain, however, we have to note that in this case there is a competing effect, since the symmetry properties of the bands also determine which are visible on top of the chain and which are visible off-side the chain. 

Based on the electron-hole oscillations of the YSR states in the vacuum observed in the simulations, %in these systems 
and the uncertainty of the absolute vertical distance of the STM tip from the surface, %we may use 
either the electron or the hole part of the calculated LDOS may provide a better qualitative agreement with the measured spectrum, as illustrated in Fig.~4 of the main text. %and usually choose the one that fits the experimental data more accurately for comparison.
%Based on these findings, we alternately use the electron or hole component of the density of state during the main text to match the theoretical results with the experiments. 

\begin{figure}
 	\centering
     \includegraphics[scale=1]{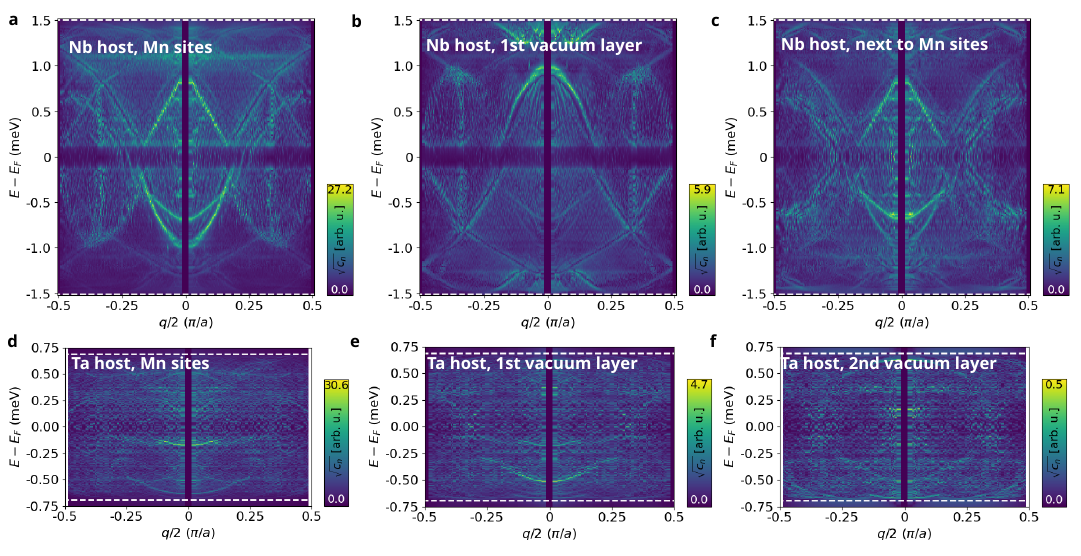}
     \caption{\label{sfig:eh-osc}
    Fourier transform of the LDOS on the Mn chains from simulations at different spatial positions. \textbf{a-c},
    Averaged 1D-FFT of the LDOS of Mn$_{L}$ chains on Nb(110) from \textit{ab initio} calculations with $10\leq L \leq36$ calculated \textbf{a}, directly on the Mn sites, \textbf{b}, in the first vacuum layer above the Mn sites, and \textbf{c}, on the vacuum sites next to the Mn chains. The magnetization points along the [110] or $z$ direction. \textbf{d-f}, Same for  Mn$_{L}$ chains on Ta(110)  with $14\leq L \leq34$ calculated \textbf{d}, directly on the Mn sites, \textbf{e}, in the first vacuum layer above the Mn sites, and \textbf{f}, in the second vacuum layer above the Mn sites. The magnetization points along the [100] or $y$ direction. White dashed lines indicate the gap of the substrate, $\Delta_{\textrm{Nb}}=1.51$~meV and $\Delta_{\textrm{Ta}}=0.69$~meV.
    %\textbf{QPI spectrum calculated along different lines around the Mn chains}\\ \textbf{a-c} The QPI spectrum of [100] Mn chains on the Ta(110) surface with magnetization along the [100] or $y$ direction. \textbf{a} Calculated on the Mn sites, \textbf{b} in the first vacuum layer above the Mn sites, \textbf{c} in the second vacuum layer above the Mn sites. \textbf{d-f} The QPI spectrum of [100] Mn chains on the Nb(110) surface with out-of-plane magnetization. \textbf{d} Calculated on the Mn sites, \textbf{e} in the first vacuum layer above the Mn sites, \textbf{f} for the vacuum sites next to the Mn chain.
    }
\end{figure}

\section{Spectral features of the odd bands on the Ta substrate}

\begin{figure}
 	\centering
     \includegraphics[scale=1]{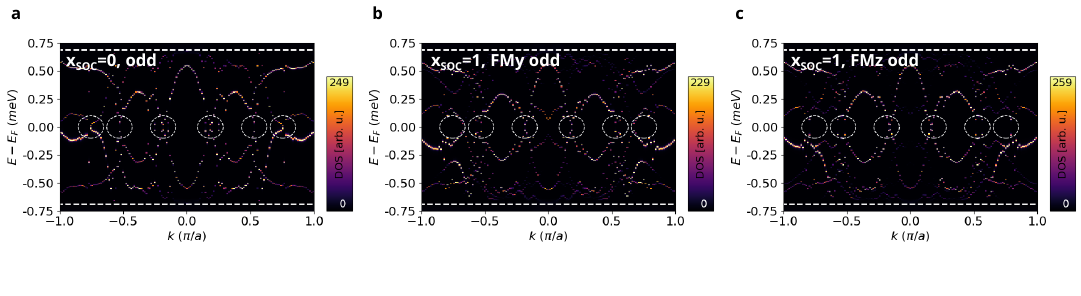}
     \caption{\label{sfig:Ta-odd}
     Spectral functions of the infinite Mn chains on Ta(110) substrate, projected to the odd orbitals. The simulations \textbf{a}, in the absence of SOC, \textbf{b}, in the presence of SOC for magnetization along the chain (FMy) and \textbf{c}, in the presence of SOC for out-of-plane magnetization (FMz) are compared. White dashed lines at $\pm\Delta_{\textrm{Ta}}=\pm0.69$~meV denote the superconducting gap of the substrate. White dashed circles highlight the regions with avoided band crossings.
    }
\end{figure}

The spectral functions of the odd bands in the infinite Mn chain on the Ta substrate are shown in Supplementary Fig.~\ref{sfig:Ta-odd}, for comparison with the even bands in Fig.~4g-i in the main text. The spectra display more features than for the Nb substrate. In the absence of SOC in Supplementary Fig.~\ref{sfig:Ta-odd}a, the data may be interpreted as displaying three Fermi-level crossings for $k>0$ at $k=0.20\frac{\pi}{a}$, $k=0.50\frac{\pi}{a}$ and $k=0.75\frac{\pi}{a}$, highlighted by white dashed circles. A few more points close to the Fermi level are visible to the left and right of the first crossing, but since they only appear at a single $k$ point and are not observable in the presence of SOC, they may be caused by numerical inaccuracies. The spectrum also approaches zero energy at the edge of the Brillouin zone, but does not cross it. Overall, this indicates an odd winding number for the even bands. In the presence of SOC for both magnetization directions, a minigap opens at $k=0.75\frac{\pi}{a}$, but no minigap can be observed at the other two crossings, where the jumps in energy as a function of wave vector may mainly be attributed to the discretization combined with the high Fermi velocities at these points. An upper limit on the possible size of the minigap may be estimated to be $\Delta_{\textrm{Ta,odd}}=0.03$~meV which is very difficult to resolve in experiments and simulations, meaning that no end states can be identified for finite chains. The hybridization with the even bands in the presence of SOC is quite pronounced, which can be observed via the appearance of a small minigap at $k=0.70\frac{\pi}{a}$ in Supplementary Fig.~\ref{sfig:Ta-odd}b (cf. the position of this avoided crossing to the leftmost and rightmost dashed circles to see that this feature is not present without SOC in Supplementary Fig.~\ref{sfig:Ta-odd}a), and of V-shaped branches close to $k=0$ between $E-E_{\textrm{F}}=0.10$~meV and $E-E_{\textrm{F}}=0.20$~meV for both magnetization directions.

\section*{Descriptions of the Supplementary Movies}
\noindent
\textbf{Supplementary Movie 1} $\vert$ Movie from the LDOS calculated for Mn chains on Nb(110) for the $x_\text{SOC}=1$ case with lengths ranging from $L=10$ to $L=37$. The dashed magenta lines mark the superconducting gap of the Nb substrate $\pm\Delta_{\mathrm{Nb}}=1.51$ meV.\\
\textbf{Supplementary Movie 2} $\vert$ Movie from the LDOS calculated for Mn chains on Nb(110) for the $x_\text{SOC}=1.25$ case with lengths ranging from $L=10$ to $L=37$. The dashed magenta lines mark the superconducting gap of the Nb substrate $\pm\Delta_{\mathrm{Nb}}=1.51$ meV.\\
\textbf{Supplementary Movie 3} $\vert$ Movie from the deconvoluted \didusig line profiles measured along the centers of the Mn chains on Ta(110) with lengths ranging from $L=2$ to $L=34$. Measurement parameters: $V_{\textrm{stab}} = -2.5$~mV, $I_{\textrm{stab}} = 1$~nA, and $V_{\textrm{mod}} = 20~\mu$V. Red and white dashed horizontal lines mark the edges of the region with reduced intensity and $\pm\Delta_{\textrm{Ta}}$, respectively. \\
\textbf{Supplementary Movie 4} $\vert$ Movie from the raw data \didusig line profiles measured along the centers of the Mn chains on Ta(110) with lengths ranging from $L=2$ to $L=34$. Measurement parameters: $V_{\textrm{stab}} = -2.5$~mV, $I_{\textrm{stab}} = 1$~nA, and $V_{\textrm{mod}} = 20~\mu$V. The gray dashed horizontal lines mark $\pm(\Delta_{\textrm{Ta}}+\Delta_{\textrm{tip}})$ with the gap of the used superconducting tip $\Delta_{\textrm{tip}}$. The energy region between $\pm\Delta_{\textrm{tip}}$ is left out for the sake of visibility.\\
\textbf{Supplementary Movie 5} $\vert$ Movie from the LDOS calculated for Mn chains, magnetized along the chain direction, on Ta(110) with lengths ranging from $L=10$ to $L=37$. The dashed magenta lines mark the superconducting gap of the Ta substrate $\pm\Delta_{\mathrm{Ta}}=0.69$ meV.\\
\textbf{Supplementary Movie 6} $\vert$ Movie from the LDOS calculated for Mn chains, magnetized out of plane, on Ta(110) with lengths ranging from $L=10$ to $L=37$. The dashed magenta lines mark the superconducting gap of the Ta substrate $\pm\Delta_{\mathrm{Ta}}=0.69$ meV.

\bibliography{references}